%% file: report.tex
\documentclass[conference]{IEEEtran}
\usepackage{amsmath,bm,qtree,cite}
\usepackage{amssymb}
\usepackage{fancyvrb}
\usepackage{multicol}
\usepackage{graphicx}
\usepackage{url}
\usepackage{tabularx}
\usepackage{caption}
\usepackage{color}
\usepackage{subcaption}
\usepackage{epstopdf}
\usepackage{boxedminipage}
\let\labelindent\relax
\usepackage{enumitem}
\usepackage{epstopdf}
\usepackage{subcaption}
\usepackage[ruled,vlined,linesnumbered]{algorithm2e}
\newtheorem{thm}{Theorem}
\newtheorem{lemma}[thm]{Lemma}

\title{Locating Contagion Sources in Networks with Partial Timestamps}
\author{\IEEEauthorblockN{Kai Zhu, Zhen Chen and Lei Ying}\\
\IEEEauthorblockA{School of Electrical, Computer and Energy Engineering\\ Arizona State University\\
Tempe, AZ, United States, 85287\\
Email: kzhu17@asu.edu, zchen113@asu.edu, lei.ying.2@asu.edu}}
\newcommand{\TruncGaussianConstant}{Z}

\newcommand{\InfNode}{{\cal I}}

\newcommand{\InfObsTimeEnt}{\tau}

\newcommand{\InfTree}{{\cal T}}
\newcommand{\InfTimeSeq}{{\bf \InfTimeSeqEnt}}
\newcommand{\InfTimeSeqEnt}{t}

\newcommand{\PropagationPath}{{\cal P}}

\newcommand{\ValidSamplePathSet}{{\cal L}}

\newcommand{\Path}{{\cal Q}}

\newcommand{\LongestPathLength}{\eta}
\newcommand{\ShortestPathLength}{l}

\begin{document}
\maketitle
\input{report/abstract}
\input{report/Introduction.tex}

\input{report/ProposedCostFunction.tex}
\input{report/ProposedAlgorithm.tex}
\input{report/ExperimentalEvaluation.tex}
\input{report/Conclusion.tex}
\bibliographystyle{IEEEtran}

\cleardoublepage
\begin{appendix}
\input{report/JustificationoftheQuadraticCostFunction.tex}

\end{appendix}
\end{document}

%% file: report/abstract.tex
\begin{abstract}
This paper studies the problem of identifying the contagion source when partial timestamps of a contagion process are available. We formulate the source localization problem as a {\em ranking problem on graphs}, where infected nodes are ranked according to their likelihood of being the source. Two ranking algorithms, cost-based ranking (CR) and tree-based ranking (TR), are proposed in this paper. Experimental evaluations with synthetic and real-world data show that our algorithms significantly improve the ranking accuracy compared with four existing algorithms.
\end{abstract} 

%% file: report/Introduction.tex
\section{Introduction}\label{sec:realworldscenario}

This paper studies the problem of identifying the contagion source when partial timestamps of a contagion process are available. Contagion processes can be used to model many real-world phenomena, including rumor spreading in online social networks, epidemics in human beings, and malware on the Internet. Informally speaking, locating the source of a contagion process refers to the problem of identifying a node in the network that provides the best explanation of the observed contagion.

This source localization problem has a wide range of applications. In epidemiology, identifying patient zero can provide important information about the disease. For example, in the Cholera outbreak in London in 1854 \cite{John_1854}, the spreading pattern of the Cholera suggested that the water pump located at the center of the spreading was likely to be the source. Later, it was confirmed that the Cholera indeed spreads via contaminated water. In online social networks, identifying the source can reveal the user who started a rumor or the user who first announced certain breaking news. For rumors, rumor source detection helps hold people accountable for their online behaviors; and for news, the news source can be used to evaluate the credibility of the news.

While locating contagion sources has these important applications in practice, the problem is difficult to solve, in particular, in complex networks. A major challenge is the lack of complete timestamp information, which prevents us from reconstructing the spreading sequence to trace back the source. But on the other hand, even partial timestamps, which are available in many practical scenarios, provide important insights about the location of the source. The focus of this paper is to develop source localization algorithms that utilize partial timestamp information.

\begin{table*}[tb]
\centering
\begin{tabular}{|c|c|c|c|c|c|c|}
  \hline
   & CR & TR & GAU & NETSLEUTH & ECCE & RUM \\
  \hline
  IAS & 0.76 & 0.68 & 0.57 & 0.43 & 0.15 & 0.15 \\
  \hline
  PG & 0.98 & 0.99 & 0.98 & 0.43 & 0.43 & 0.39\\
  \hline
\end{tabular}
  \caption{The $10\%$-accuracy under different source localization algorithms with 50\% timestamps}
  \label{tab:intro}
\end{table*}

We remark that while this source localization problem (or called rumor source detection problem) has been studied recently under a number of different models (see Section \ref{sec:related} for the details), most of them ignore timestamp information. As we will see from the experimental evaluations, even limited timestamp information can significantly improve the accuracy of locating the source. In this paper, we use a {\em ranking-on-graphs} approach to exploit the timestamp information, and develop source localization algorithms that perform well on different networks and under different contagion models. The main contributions of this paper are summarized below.
\begin{enumerate}[leftmargin=*]
\item[(1)] We formulate the source localization problem as a {ranking problem on graphs}, where infected nodes are ranked according to their likelihood of being the source. Define a {\em spreading tree} to include (i) a directed tree with all infected nodes and (ii) the complete timestamps of contagion propagation (the detailed definition will be presented in Section \ref{sec:approach}). Given a spreading tree rooted at node $v,$ denoted by ${\cal P}_v,$ we define a quadratic cost $C({\cal P}_v)$ depending on the structure of the tree and the timestamps. The cost of node $v$ is then defined to be \begin{equation}C(v)=\min_{{\cal P}_v} C({\cal P}_v),\label{eq:cost}\end{equation} i.e., the minimum cost among all spreading trees rooted at node $v.$ Based on the costs and spreading trees, we propose two ranking methods:
    \begin{itemize}
    \item[(i)] rank the infected nodes in an ascendent order according to $C(v)$, called {\em cost-based ranking (CR)}, and
    \item[(ii)] find the minimum cost spreading tree, i.e., $${\cal P}^*=\arg\min_{{\cal P}} C({\cal P}),$$ and rank the infected nodes according to their timestamps on the minimum cost spreading tree, called {\em tree-based ranking (TR)}.

    \end{itemize}

\item[(2)] The computational complexity of $C(v)$ is very high due to the large number of possible spreading trees. We prove that problem (\ref{eq:cost}) is NP-hard by connecting it to the longest-path problem \cite{GarJoh_79}.

\item[(3)] We propose a greedy algorithm, named Earliest Infection First (EIF), to construct a spreading tree to approximate the minimum cost spreading tree for a given root node $v,$ denoted by $\tilde{\cal P}_v.$ The greedy algorithm is designed based on the minimum cost solution for line networks. EIF first sorts the infected nodes with observed timestamps in an ascendent order of the timestamps, and then iteratively attaches these nodes using a modified breadth-first search algorithm. In CR, the infected nodes are then ranked based on $C(\tilde{\cal P}_v);$ and in TR, the nodes are ranked based on the complete timestamps of the spreading tree $\tilde{\cal P}^*$ such that $$\tilde{\cal P}^*=\arg\min C(\tilde{\cal P}_v).$$ We remark that for infected nodes with unknown infection time, EIF assigns the infection timestamps during the construction of the spreading tree $\tilde{\cal P}_v.$ The details can be found in Section \ref{sec:alg}.


\item[(4)] We conducted extensive experimental evaluations using both synthetic data and real-world social network data (Sina Weibo\footnote{\url{http://www.weibo.com/}}). The performance metric is the probability with which the source is ranked among top $\gamma$ percent, named $\gamma\%$-accuracy. We have the following observations from the experimental results:
     \begin{itemize}
     \item [(i)] Both CR and TR significantly outperform existing source location algorithms in both synthetic data and real-world data. Table \ref{tab:intro} summarizes the $10\%$-accuracy in the Internet autonomous systems (IAS) network and the power grid (PG) network.

     \item[(ii)] Our results show that both TR and CR perform well under different {contagion models} and different {distributions of timestamps.}

    \item[(iii)] Early timestamps are more valuable for locating the source than recent ones.

     \item[(iv)] Network topology has a significant impact on the performance of source localization algorithms, including both ours and existing ones. For example, the $\gamma\%$-accuracy in the IAS network is lower than that in the PG network (see Table \ref{tab:intro} for the comparison). This suggests that the problem is more difficult in networks with small diameters and hubs than in networks that are locally tree-like.
     \end{itemize}
\end{enumerate}

\subsection{Related Work}
\label{sec:related}
Contagion source detection has received a lot of attention recently. In \cite{ShaZam_11,ShaZam_12}, Shah and Zaman developed rumor centrality under the susceptible-infected (SI) model, and proved that rumor centrality is the maximum likelihood estimator for regular trees. The use of rumor centrality for source detection has been extended to other scenarios including multiple sources \cite{LuoTayLen_13}, single source with partial observations \cite{KarFra_13}, single source with a priori distribution \cite{DonZhaTan_13}, and single source with multiple infection instants \cite{WanDonZha_14}. In \cite{ZhuYin_13}, Zhu and Ying developed a sample-path-based approach for detecting the source under the susceptible-infected-recovered (SIR) model. The sample-path-based approach has been applied to the SI model with partial observations  \cite{LuoTay_13_2}, SIR model with partial observations \cite{ZhuYin_14}, SIS model \cite{LuoTay_13}, and  SIR model with multiple sources \cite{CheZhuYin_14}. Other solutions in the literature include NETSLEUTH, an eigenvalue-based estimator  \cite{PraVreFal_12}, a dynamic message-passing algorithm \cite{LokMezOht_13} and a minimum-cost-based solution for the SI model with sparse observations \cite{GunFenLiu_2013}. None of these algorithms utilize the timestamp information, which is the focus of this paper.

The work mostly related to ours is  \cite{PinThiVet_12}, which leverages timestamp information from observers to locate the source. The algorithm is similar to CR in spirit, but uses the breadth-first search tree as the spreading tree from a given infected node. In the IAS network, not only the performance of the algorithm is worse than ours, the gap also increases significantly as the amount of timestamps increases (e.g., with 90\% timestamps, the $10\%$-accuracy of CR is 1 and that of the algorithm in \cite{PinThiVet_12} is only 0.5). We conjecture this is because the spreading trees constructed by EIF (our algorithm) is far more accurate than the breadth-first search trees.

Another algorithm that exploits timestamp information is the simulation-based fast Monte Carlo algorithm \cite{AgaLu_13}, which utilizes multiple snapshots from a set of sparsely placed nodes. The approach, however, requires the infection time distributions of all edges, which are difficult to obtain in practice.

%% file: report/ProposedCostFunction.tex
\section{A Ranking Approach for Source Localization}
\label{sec:approach}
Ideally, the output of a source localization algorithm should be a single node, which matches the source with a high probability. However, with limited timestamp information, this goal is too ambitious, if not impossible, to achieve. From the best of our knowledge, almost all evaluations using real-world networks show that the detection rates of existing source localization algorithms are very low \cite{ShaZam_11,ZhuYin_13,CheZhuYin_14,LuoTayLen_13}, where the detection rate is the probability that the detected node is the source.

When the detection rate is low, instead of providing a single source estimator, a better and more useful output of a source localization algorithm would be a node ranking, where nodes are ordered according to their likelihood of being the source.  With such a ranking, further investigation can be conducted to locate the source. More accurate the ranking, less amount of resources is needed in the further investigation.

\begin{figure*}
        \centering
        \begin{subfigure}[b]{0.48\textwidth}
                \centering
                 \includegraphics[width=\textwidth]{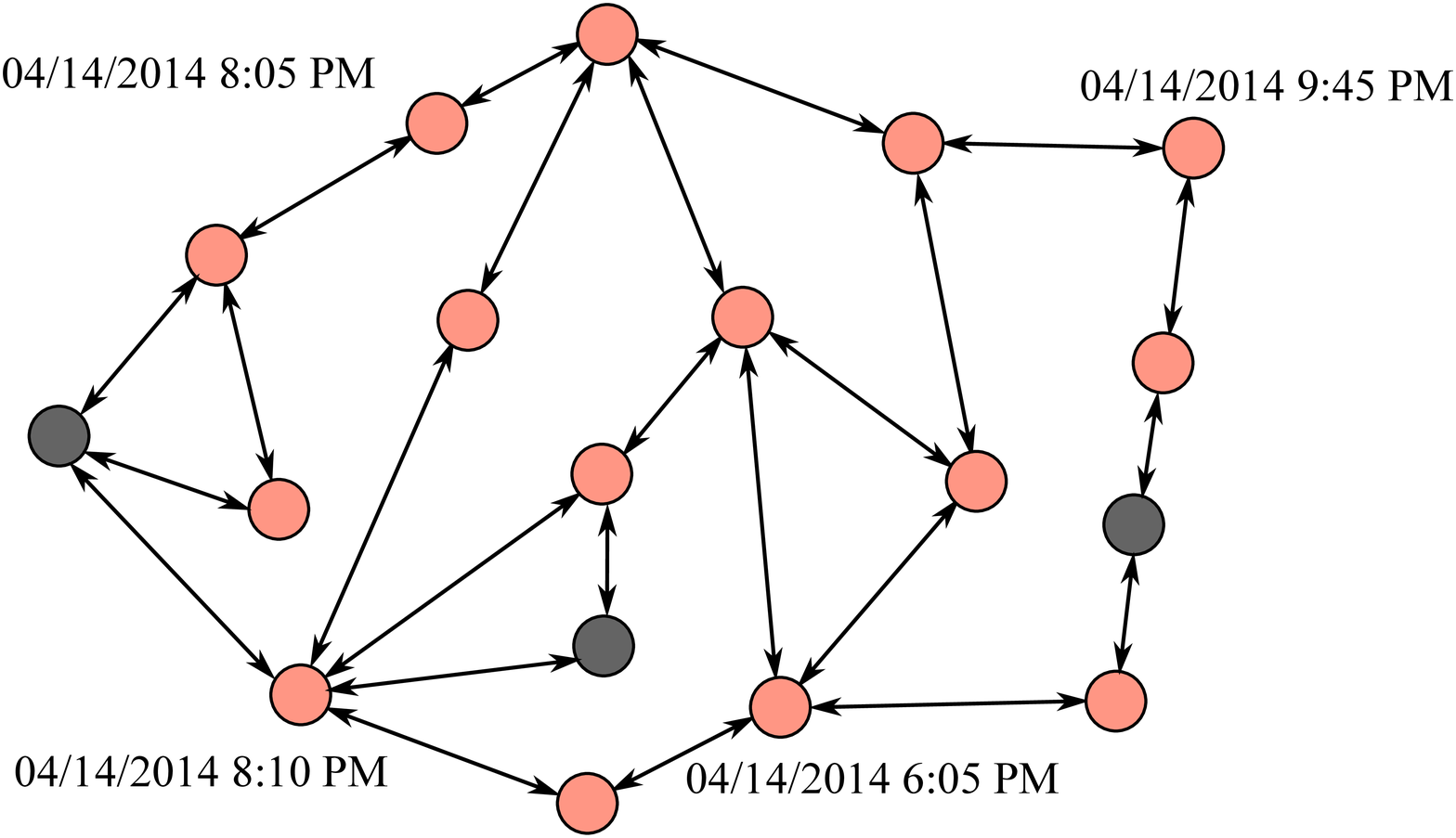}
                  \caption{Available Partial Timestamps}\label{figure:RealWorldExample}
        \end{subfigure}
~
        \begin{subfigure}[b]{0.48\textwidth}
                \centering
                  \includegraphics[width=\textwidth]{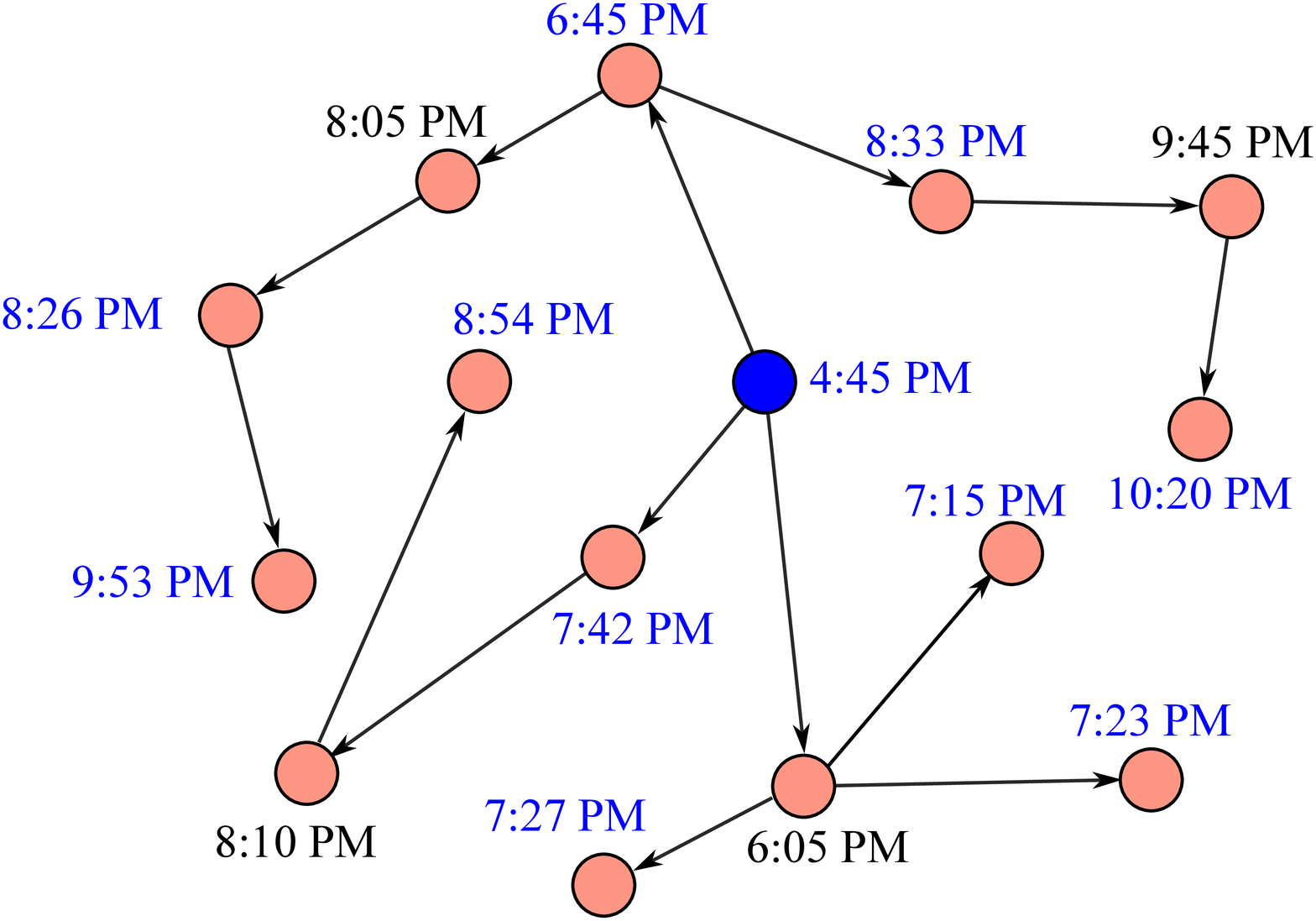}
                  \caption{An Feasible and Consistent Spreading Tree}\label{figure:FeasibleAndConsistentExample}
        \end{subfigure}
        \centering
        \caption{An Example Illustrating Available Information and a Spreading Tree}
\end{figure*}

In this paper, we assume the input of a source localization algorithm includes the following information:
\begin{itemize}
\item {\em A network $G({\cal V}, {\cal E})$:} The network is an unweighted and directed graph. A node $v$ in the network represents a physical entity (such as a user of an online social network, a human being, or a mobile device). A directed edge $e(v, u)$ from node $v$ to node $u$ indicates that the contagion can be transmitted from node $v$ to node $u.$

\item {\em A set of infected nodes ${\cal I}$:} An infected node is a node that involves in the contagion process, e.g., a twitter user who retweeted a specific tweet, a computer infected by malware, etc. We assume $\cal I$ includes all infected nodes in the contagion. So $\cal I$ forms a connected subgraph of $G.$ In the case $\cal I$ includes only a subset of infected nodes, our source localization algorithms rank the observed infected nodes according to their likelihood of being the earliest infected node. More discussion can be found in Section \ref{sec:con}.

\item {\em Partial timestamps $\bm \tau$:} $\bm \tau$ is a $|{\cal V}|$-dimensional vector such that $\tau_v=\star$ if the timestamp is missing and otherwise, $\tau_v$ is the time at which node $v$ was infected. We remark that the time here is {\em the normal clock time, not the relative time with respect to the infection time of the source.} Note that in most cases, the infection time of the source is as difficult to know as the location of the source.

\end{itemize}

Figure \ref{figure:RealWorldExample} is a simple example showing the available information. The nodes in orange are the infected nodes. The time next to a node is the associated timestamp. We define a spreading tree $\PropagationPath=(\InfTree,\InfTimeSeq)$ to be a directed tree $\InfTree$ with a $|\InfTree|$-dimensional vector $\InfTimeSeq.$ The directed tree $\InfTree$ specifies the sequence of infection and the vector $\InfTimeSeq$ specifies the time at which each infection occurs. We further require the time sequence $\bf t$ of a spreading tree to be \emph{feasible} such that the infection time of a node is larger than its parent's, and to be \emph{consistent} with the partial timestamps $\bm \tau$ such that $t_v=\tau_v$ if $\tau_v\not=\star.$  Figure \ref{figure:FeasibleAndConsistentExample} shows a spreading tree that is feasible and  consistent with the observation shown in Figure \ref{figure:RealWorldExample}. Note that, for simplicity, we omitted the date in the figure by assuming all events occur on the same day. The timestamps in black are the observed timestamps and the ones in blue are assigned by us. Denote by $\ValidSamplePathSet({\cal I}, {\bm \tau})$ the set of spreading trees that are both feasible and consistent with the partial timestamps.

\subsection{Quadratic cost and sample path approach}
Given a spreading tree $\PropagationPath=(\InfTree,\InfTimeSeq) \in\ValidSamplePathSet({\cal I}, {\bm \tau}),$ we define the cost of the tree to be
 \begin{equation}
 C(\PropagationPath) = \sum_{(v,w)\in \InfTree}(\InfTimeSeqEnt_w-\InfTimeSeqEnt_v-\mu)^2, \label{eq:q-cost}
 \end{equation}
 for some constant $\mu>0.$ This quadratic cost function is motivated by the following model.

For each edge $(v,w)\in {\cal T},$ assume that the time it takes for node $v$ to infect node $w$ follows a truncated Gaussian distribution with mean $\mu$ and variance $\sigma^2.$ Then given a spreading tree $\PropagationPath,$ the probability density associated with time sequence $\InfTimeSeq$ is
 \begin{equation}
 f_\PropagationPath(\InfTimeSeq)=\prod_{(v,w)\in \InfTree}\frac{1}{\TruncGaussianConstant\sqrt{2\pi}\sigma}\exp\left(-\frac{(\InfTimeSeqEnt_w-\InfTimeSeqEnt_v-\mu)^2}{2\sigma^2}\right),\label{eqn:pdfOfPropagationPath}
 \end{equation}
 where $\TruncGaussianConstant$ is the normalization constant. Therefore, the log-likelihood is
\begin{align*}
&\log f_\PropagationPath(\InfTimeSeq)\\
=&-|{\cal E}(\InfTree)|\log(\TruncGaussianConstant\sqrt{2\pi}\sigma)-\frac{1}{2\sigma^2}\sum_{(v,w)\in \InfTree}(\InfTimeSeqEnt_w-\InfTimeSeqEnt_v-\mu)^2,
\end{align*} where $|{\cal E}(\InfTree)|$ is the number of edges in the tree. 
Therefore, given a tree $\InfTree,$ the log-likelihood of time sequence  $\InfTimeSeq$ is inversely proportional to  the quadratic cost defined in (\ref{eq:q-cost}). The lower the cost, more likely the time sequence occurs.

Now given an infected node in the network, the cost of the node is defined to be minimum cost among all spreading trees rooted at the node. Using ${\cal P}_v$ to denote a spreading tree rooted at node $v,$ the cost of node $v$ is
  \begin{align}
  C(v) = \min_{\PropagationPath_v\in \ValidSamplePathSet({\cal I}, {\bm \tau})} C(\PropagationPath_v). \label{eqn:cost of nodes}
\end{align}

After obtaining $C(v)$ for each infected node $v,$  the infected nodes can be ranked according to either $C(v)$ or the timestamps of the minimum cost spreading tree. However, the calculation of $C(v)$ in a general graph is NP-hard as shown in the following theorem.
\begin{thm}
Problem (\ref{eqn:cost of nodes}) is an NP-hard problem. \label{thm:np}\hfill{$\square$}
\end{thm}
{\bf Remark 1:} This theorem is proved by showing that the longest-path problem can be solved by solving (\ref{eqn:cost of nodes}). The detailed analysis is presented in the appendix. Since computing the exact value of $C(v)$ is difficult, we present a greedy algorithm in the next section.

%% file: report/ProposedAlgorithm.tex
\section{EIF: A Greedy Algorithm}
\label{sec:alg}
\begin{figure*}[tbh]
\begin{centering}
  \includegraphics[width=\textwidth]{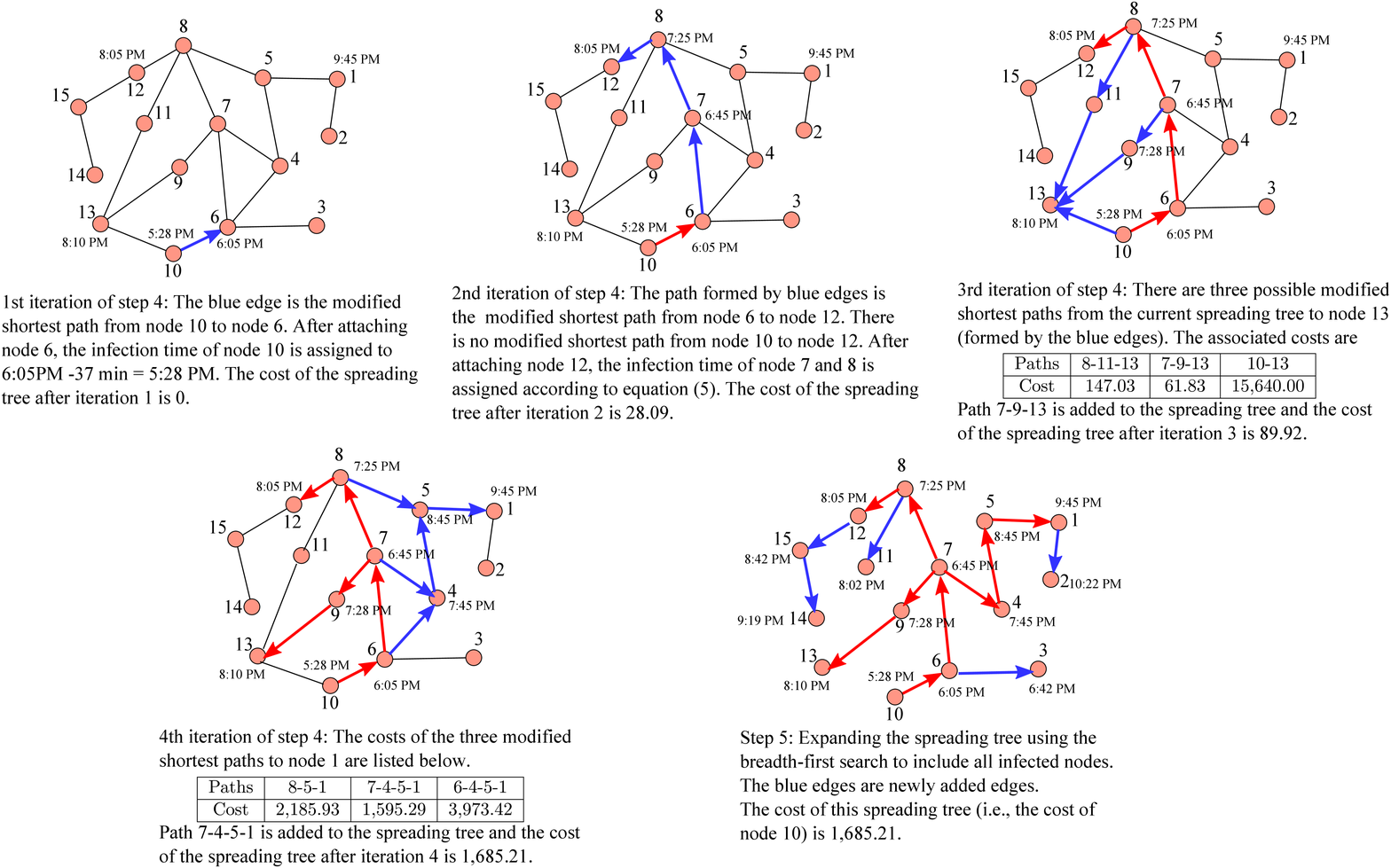}
  \caption{An Example for Illustrating Step 4 and Step 5 of EIF. The paths formed by blue edges are modified shortest paths. The trees formed by red edges are the spreading trees at the beginning of each iteration. }\label{figure:AlgorithmExecutionExample}
  \end{centering}
\end{figure*}

In this section, we present a greedy algorithm, named Earliest-Infection-First (EIF), to solve problem (\ref{eqn:cost of nodes}). Note that if a node's observed infection time is larger than some other node's observed infection time, then it cannot be the source. So we only need to compute cost $C(v)$ for node $v$ such that $\tau_v=\star$ or $\tau_v=\min_{u: \tau_u\not=\star} \tau_u.$ Furthermore, when all infected nodes are known, we can restrict the network to the subnetwork formed by the infected nodes to run the algorithm. We next present the algorithm, together with a simple example in Figure \ref{figure:AlgorithmExecutionExample} for illustration.  In the example, all edges are {\em bidirectional,} so the arrows are omitted, and the network in Figure \ref{figure:AlgorithmExecutionExample} is the subnetwork formed by all infected nodes.

\vspace{0.1in}
\hrule
\vspace{0.1in}

\noindent{\bf Earliest-Infection-First (EIF)}

\begin{enumerate}[leftmargin=*]

\item Step 1: The algorithm first estimates $\mu$ from $\bm \tau$ using the average per-hop infection time. Let $l_{vw}$ denote the length of the shortest path from node $v$ to node $w,$ then
$$\mu=\frac{\sum_{\tau_v\not=\star,\tau_w\not=\star, v\neq w}|\InfObsTimeEnt_v-\InfObsTimeEnt_w|}{\sum_{\tau_v\not=\star,\tau_w\not=\star, v\neq w}\ShortestPathLength_{vw}}.$$

{\bf Example:} Given the timestamps shown in Figure \ref{figure:AlgorithmExecutionExample}, $\mu=36.94 \hbox{ minutes}.$

\item Step 2: Sort the infected nodes in an ascending order according to the observed infection time $\bm \tau.$ Let $\alpha$ denote the ordered list such that $\alpha_1$ is the node with the earliest infection time.

    {\bf Example:} Consider the example in Figure \ref{figure:AlgorithmExecutionExample}. The ordered list is $$\alpha=(6, 12, 13, 1).$$

\item Step 3: Construct the initial spreading tree $\InfTree_0$ that includes the root node only and set the cost to be zero.

{\bf Example:} Assuming we want to compute the cost of node 10 in Figure \ref{figure:AlgorithmExecutionExample}, we first have $\InfTree_0=\{10\}$ and $C(10)=0.$

\item Step 4: At the $k^{\rm th}$ iteration, node $\alpha_k$ is added to the spreading tree $\InfTree_{k-1}$ using the following steps.

    {\bf Example:} At $3^{\rm rd}$ iteration, the current spreading tree is $$10\rightarrow 6 \rightarrow 7 \rightarrow 8 \rightarrow 12,$$ and the associated timestamps are given in Table \ref{tab:time}. Note that these timestamps are assigned by EIF except those observed ones. The details can be found in the next step.  In the $3^{\rm rd}$ iteration, node 13 needs to be added to the spreading tree.

\begin{table}[htb]
\centering
\begin{tabular}{|c|c|c|c|c|c|}
  \hline
  node ID & 10 & 6 & 7 & 8 & 12 \\
  \hline
  Timestamp & 5:28 & 6:05 & 6:45  & 7:25  & 8:05 \\
  \hline
\end{tabular}
\caption{The timestamps on the spreading tree in the $3^{\rm rd}$ iteration} \label{tab:time}
\end{table}

\begin{itemize}
\item[(a)] For each node $m$ on the spreading tree $\InfTree_{k-1},$ identify a modified shortest path from node $m$ to node $\alpha_k.$ The modified shortest path is a path that has the minimum number of hops among all paths from node $m$ to node $\alpha_k,$ which satisfy the following two conditions:
    \begin{itemize}
    \item it does not include any nodes on the spreading tree $\InfTree_{k-1},$ except node $m;$ and
    \item it does not include any nodes on list $\alpha,$ except node $\alpha_k.$
    \end{itemize}

{\bf Example:} The modified shortest path from node $7$ to node $13$ is $$7\rightarrow 9 \rightarrow 13.$$ There is no modified shortest path from node $12$ to node $13$ since all paths from $12$ to $13$ go through node $8$ that is on the spreading tree $\InfTree_{2}$.

\item[(b)] For the modified shortest path from node $m$ to node $\alpha_k,$ the cost of the path is defined to be $$\gamma_m=\tilde{l}_{\alpha_km}\left(\frac{t_{\alpha_k}-t_m}{\tilde{l}_{\alpha_km}}-\mu\right)^2,$$
where $\tilde{l}_{\alpha_km}$ denotes the length of the modified shortest path from $m$ to $\alpha_k.$  From all nodes on the spreading tree ${\cal T}_{k-1},$ select node $m^*$ with the minimum cost i.e., $$m^*=\arg\min_m \gamma_m.$$

{\bf Example:} The costs of the modified shortest paths to the nodes  on the spreading tree $$10\rightarrow 6 \rightarrow 7 \rightarrow 8 \rightarrow 12$$ are shown in Table \ref{tab:cost}. Node 7 has the smallest cost.
\begin{table}[htb]
\centering
\begin{tabular}{|c|c|c|c|c|c|}
  \hline
  node ID & 10 & 6 & 7 & 8 & 12 \\
  \hline
  cost & 15,640.00 & $\infty$ & 61.83 & 147.03 & $\infty$ \\
  \hline
\end{tabular}
\caption{The costs of the modified shortest paths} \label{tab:cost}
\end{table}

\item[(c)] Construct a new spreading tree ${\cal T}_k$ by adding the modified shortest path from $m^*$ to $\alpha_k.$ Assume node $g$ on the newly added path is $h_g$ hops from node $m^*,$ the infection time of node $g$ is set to be \begin{equation}t_g=t_{m^*}+(h_g-1)\frac{t_{\alpha_k}-t_{m^*}}{\tilde{l}_{m^*\alpha_k}}.\label{eq: assigntime}\end{equation} The cost is updated to $C(v)=C(v)+\gamma_{m^*}.$

{\bf Example:} At the $3^{\rm rd}$ iteration, the timestamp of node $9$ is set to be 7:28 PM, and the cost is updated to $C(10)=89.92.$
\end{itemize}

\item Step 5: For those infected nodes that have not been added to the spreading tree, add these nodes by using a breadth-first search starting from the spreading tree ${\cal T}.$ When a new node (say node $w$) is added to the spreading tree during the breadth-first search, the infection time of the node is set to be $t_{p_w}+\mu,$ where $p_w$ is the parent of node $w$ on the spreading tree. Note that the cost $C(v)$ does not change during this step because $t_w-t_{p_w}-\mu=0.$

{\bf Example:} The final spreading tree and the associated timestamps are presented in Figure \ref{figure:AlgorithmExecutionExample}.
\end{enumerate}

\vspace{0.1in}
\hrule
\vspace{0.1in}


{\bf Remark 2:} The timestamps of nodes on a newly added path are assigned according to equation (\ref{eq: assigntime}). This is because such an assignment is the minimum cost assignment in a line network in which only the timestamps of two end nodes are known.

\begin{lemma}\label{lem:assigntime}
Consider a line network with $n$ infected nodes. Assume the infection time of node $1$ and node $n$ is known and the infection time of the rest nodes is not. Furthermore, assume $\InfObsTimeEnt_1<\InfObsTimeEnt_n.$ The quadratic cost defined in (\ref{eqn:cost of nodes}) is minimized by setting
\begin{align}
\InfTimeSeqEnt_k=\InfObsTimeEnt_1+(k-1)\frac{\InfObsTimeEnt_n-\InfObsTimeEnt_1}{n-1}\label{eqn:LineTime}
\end{align}
for $1<k<n.$ \hfill{$\square$}
\end{lemma}
Note that under the assignment above, the infection time, $\tau_{k+1}-\tau_k,$ is the same for all edges, which is due to the quadratic form of the cost function. The detailed proof can be found in the appendix.

{\bf Remark 3:} Note that in Step 4(a), we use the modified shortest path instead of the conventional shortest path. The purpose is to avoid inconsistence when assigning timestamps. For example, consider the $3^{\rm rd}$ iteration in Figure \ref{figure:AlgorithmExecutionExample}, and the paths from node $7$ to node $2.$ There are two conventional shortest paths: $7\rightarrow 4\rightarrow 5\rightarrow 1$ and $7\rightarrow 8 \rightarrow 5\rightarrow 1.$  If we select path $7\rightarrow 8 \rightarrow 5\rightarrow 1$ and assign the timestamps according to (\ref{eq: assigntime}), then the infection time of node $8$ is larger that of node $7,$ which contradicts the current timestamps of node $7$ and node $8.$ Therefore, $7\rightarrow 8 \rightarrow 5\rightarrow 1$ should not be selected.

{\bf Remark 4:} A key step of EIF is the construction of the modified shortest paths from the nodes on ${\cal T}_{k-1}$ to  node $\alpha_k.$ This can be done by constructing a modified breadth-first search tree starting from node $\alpha_k.$ In constructing the modified breadth-first search tree, we first reverse the direction of all edges as we want to construct paths from the nodes on ${\cal T}_{k-1}$ to node $\alpha_k.$ Then starting from node $\alpha_k,$ nodes are added in a breadth-first fashion. However, a branch of the tree terminates when the tree meets a node on ${\cal T}_{k-1}$ or node $\alpha_l$ for $l>k.$ After obtainng the modified breadth-first search tree, if a leaf node is a node on ${\cal T}_{k-1},$ say node $m,$ then the {\em reversed path} from node $\alpha_k$ to node $m$ on the modified breadth-first search tree is a modified shortest path from node $m$ to node $\alpha_k.$ If none of the leaf nodes is on ${\cal T}_{k-1},$ then the cost of adding $\alpha_k$ is claimed to be infinity. In Figure \ref{figure:AlgorithmExecutionExample}, the trees formed by the blue edges are the modified breadth-first trees at each iteration.

\section{Cost-Based and Tree-Based Ranking}

Denote by $\tilde{\cal T}_v$ the spreading tree constructed under EIF for node $v,$ and $\tilde{C}(\tilde{\cal T}_v)$ the corresponding cost computed by EIF. After constructing the spreading tree for each infected node and obtaining the corresponding cost, the nodes are ranked using the following two approaches.

\vspace{0.1in}
\hrule
\vspace{0.1in}
\noindent{\bf Cost-Based Ranking (CR):} Rank the infected nodes in an ascendent order according to $\tilde{C}(\tilde{\cal T}_v).$

\vspace{0.1in}
\hrule
\vspace{0.1in}


\vspace{0.1in}
\hrule
\vspace{0.1in}

\noindent{\bf Tree-Based Ranking (TR):} Denote by ${v^*}=\arg\min_v \tilde{C}(\tilde{\cal T}_v).$ Rank the infected nodes in an ascendent order according to the timestamps on $\tilde{\cal T}_{v^*}.$

\vspace{0.1in}
\hrule
\vspace{0.1in}

\begin{thm}
The complexity of CR and TR is $O(|\alpha||{\cal I}||{\cal E}_I|),$ where $|\alpha|$ is the number of infected nodes with observed timestamps, $|{\cal I}|$ is the number of infected nodes, and $|{\cal E}_I|$ is the number of edges in the subgraph formed by the infected nodes. \label{thm:complexity} \hfill{$\square$}
\end{thm}

The proof is presented in the appendix. 

%% file: report/ExperimentalEvaluation.tex
\section{Experimental Evaluation}
In this section, we evaluate the performance of TR and CR using both synthetic data and real-world data.
\color{black}
\subsection{Performance of EIF on a Small Network}
In the first set of simulations, we evaluated the performance of EIF of solving the minimum cost of the feasible and consistent spreading trees. Given an observation ${\cal I}$ and ${\bm \tau},$ denote by $C^*$  the minimum cost of the feasible and consistent spreading trees. Then
\[
C^*=\min_{\PropagationPath\in \ValidSamplePathSet({\cal I}, {\bm \tau})} C(\PropagationPath)
\]
Denote by $\tilde{C}^*$ the minimum cost of the spreading trees obtained under EIF. We evaluated the approximation ratio $r=\frac{\tilde{C}^*}{C^*}$ on a small network --- the Florentine families network\cite{BrePhi_1986} which has $15$ nodes and $20$ edges. Recall that the minimum cost problem is NP-hard, so the approximation ratio is evaluated over a small network only. To compute the actual minimum cost, we first enumerated all possible spanning trees using the algorithm in \cite{Cha_1968}, and then computed the minimum cost of each spanning tree by solving the quadratic programming problem. 

\begin{figure}[htb]
\begin{centering}
  \includegraphics[width=0.4\textwidth]{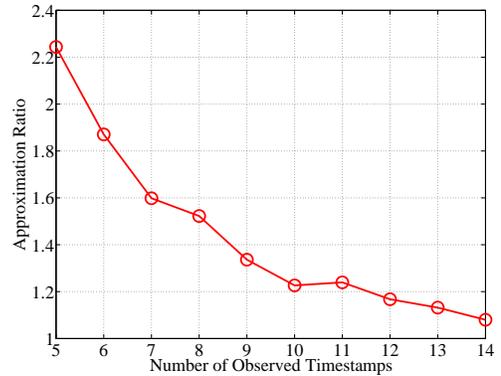}
  \caption{The Approximation Ratio of TR}\label{figure:ApproximateRatio}
 \end{centering}
\end{figure}

In this experiment, we assumed the infection time of each edge follows a truncated Gaussian distribution with $\mu=100$ and $\sigma=100.$ We evaluated the approximation ratio when the number of observed timestamps varied from 5 to 14. The results are shown in Figure \ref{figure:ApproximateRatio}, where each data point is an average of 500 runs. The approximation ratio is 2.24 with 5 timestamps, 1.5 with 8 timestamps and becomes 1.08 when 14 timestamps are given. This experiment shows that EIF approximates the minimum cost solution reasonably well.
\color{black}
\subsection{Comparison with Other Algorithms}\label{sec:comGraphCentrality}

We first tested the algorithms using synthetic data on two real-world networks: the Internet Autonomous Systems network (IAS)\footnote{Available at
\url{http://snap.stanford.edu/data/index.html}} and the power
grid network (PG)\footnote{Available at
\url{http://www-personal.umich.edu/~mejn/netdata/}}:
\begin{itemize}
\item The IAS network is a network of the Internet autonomous systems inferred from Oregon route-views on March, 31st, 2001. The network contains 10,670 nodes and 22,002 edges in the network. IAS is a small world network.

\item The PG network is {\color{black} a network of Western States Power Grid of United States.} The network contains 4,941 nodes and 6,594 edges. Compared to the IAS network, the PG network is locally tree-like.
\end{itemize}

We first compare CR and TR with the following four existing source localization algorithms.
\begin{itemize}
\item Rumor centrality (RUM): Rumor centrality was proposed in \cite{ShaZam_11}, and is the the maximum likelihood estimator on trees under the SI model. RUM ranks the infected nodes in an ascendent order according to nodes' rumor centrality.

\item Infection eccentricity (ECCE): The infection eccentricity of a node is the maximum distance from the node to any infected node in the graph, where the distance is defined to be the length of the shortest path. The node with the smallest infection eccentricity, named Jordan infection center, is the optimal sample-path-based estimator on tree networks under the SIR model \cite{ZhuYin_13}. ECCE ranks the infected nodes in a descendent order according to infection eccentricity. 

\item {\color{black} NETSLEUTH: NETSLEUTH was proposed in \cite{PraVreFal_12}. The algorithm constructs a submatrix of the infected nodes based on the graph Laplacian of the network and then ranks the infected nodes according to the eigenvector corresponding to the largest eigenvalue of the submatrix.}

\item \textcolor{black}{Gaussian heuristic (GAU): Gaussian heuristic is an algorithm proposed in \cite{PinThiVet_12}, which utilizes partial timestamp information. The algorithm is similar to CR in spirit, but uses the breadth-first search tree as the spreading tree for each infected node. }
\end{itemize}
In the four algorithms above, RUM, ECCE, and NETSLEUTH only use topological information of the network, and do not exploit the timestamp information. GAU utilizes partial timestamp information.

In this set of experiments, we assume the infection time of each infection follows a truncated Gaussian distribution with $\mu=\{1,10,100\}$ and $\sigma=100.$ \textcolor{black}{ In each simulation, a source node was chosen uniformly across node degree to avoid the bias towards small degree nodes (In the IAS network, 3,720 out of the 10,670 nodes have degree one).  In particular, the nodes were grouped into $M$ bins such that the nodes in the $m^{\rm th}$ bin ($1\leq m\leq M-1$) have degree $m$ and the nodes in the $M^{\rm th}$ bin have degree $\geq M$. In each simulation, we first randomly and uniformly picked a bin, and then randomly and uniformly pick a node from the selected bin. We simulated the contagion process and terminated the process when having 200 infected nodes. For the IAS network, we chose $M=20;$ and for the PG network, we chose $M=10.$} Note that we had more bins in the IAS network than in the PG network because the degree distribution of the IAS network is heavily tailed and the maximum degree is 2,312. The maximum degree of the PG network is only 19.

We selected 50\% infected nodes (100 nodes) and revealed their infection time. The source node was always excluded from these 100 nodes so that the infection time of the source node was always unknown. We repeated the simulation 500 times to compute the average $\gamma\%$-accuracy.



\begin{figure*}[!t]
        \centering
        \begin{subfigure}[b]{0.32\textwidth}
                \centering
                \includegraphics[width=\textwidth]{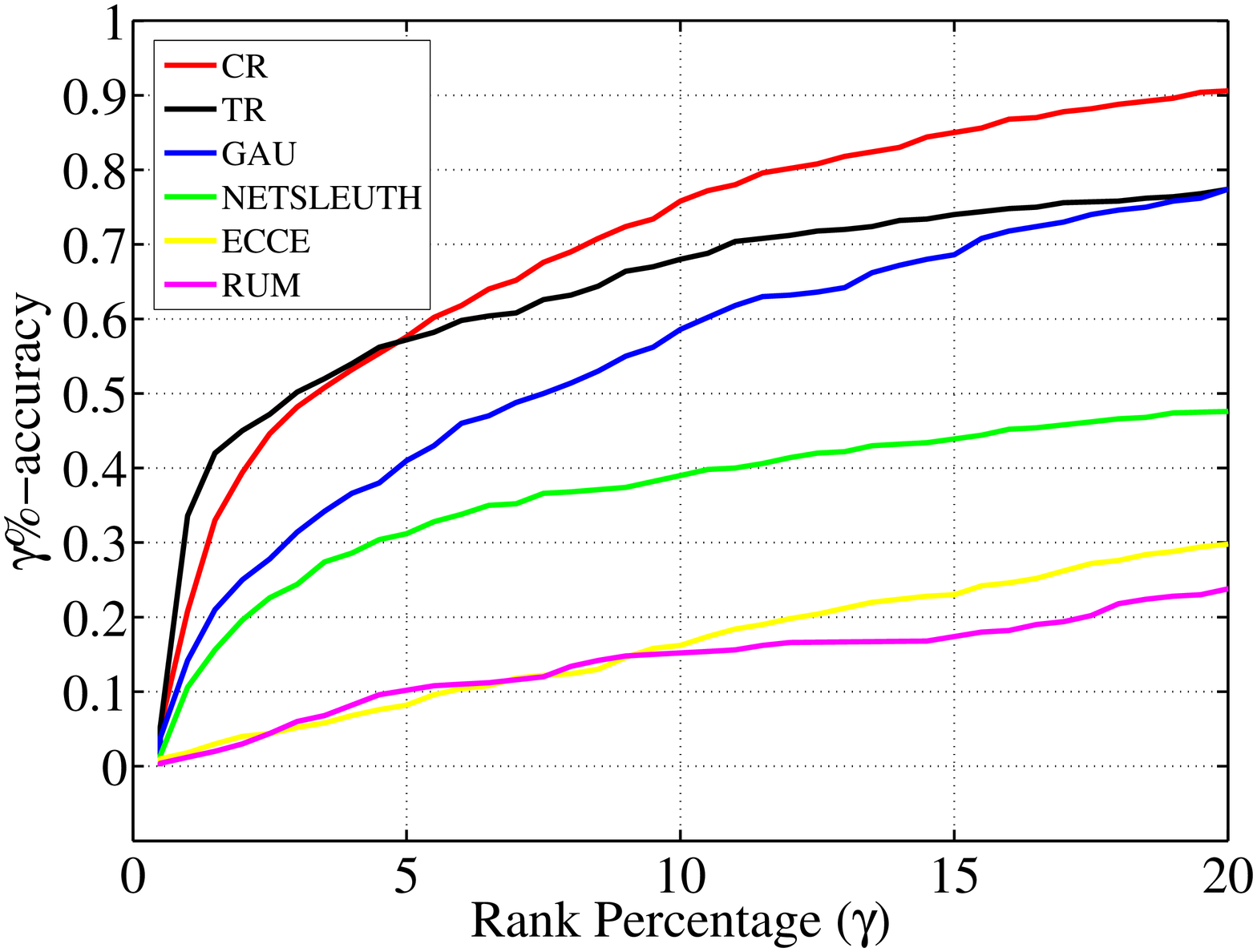}
                  \caption{The IAS network with $\mu=1$}\label{figure:CompCentralityIASmu1}
        \end{subfigure}
~
        \begin{subfigure}[b]{0.32\textwidth}
                \centering
                  \includegraphics[width=\textwidth]{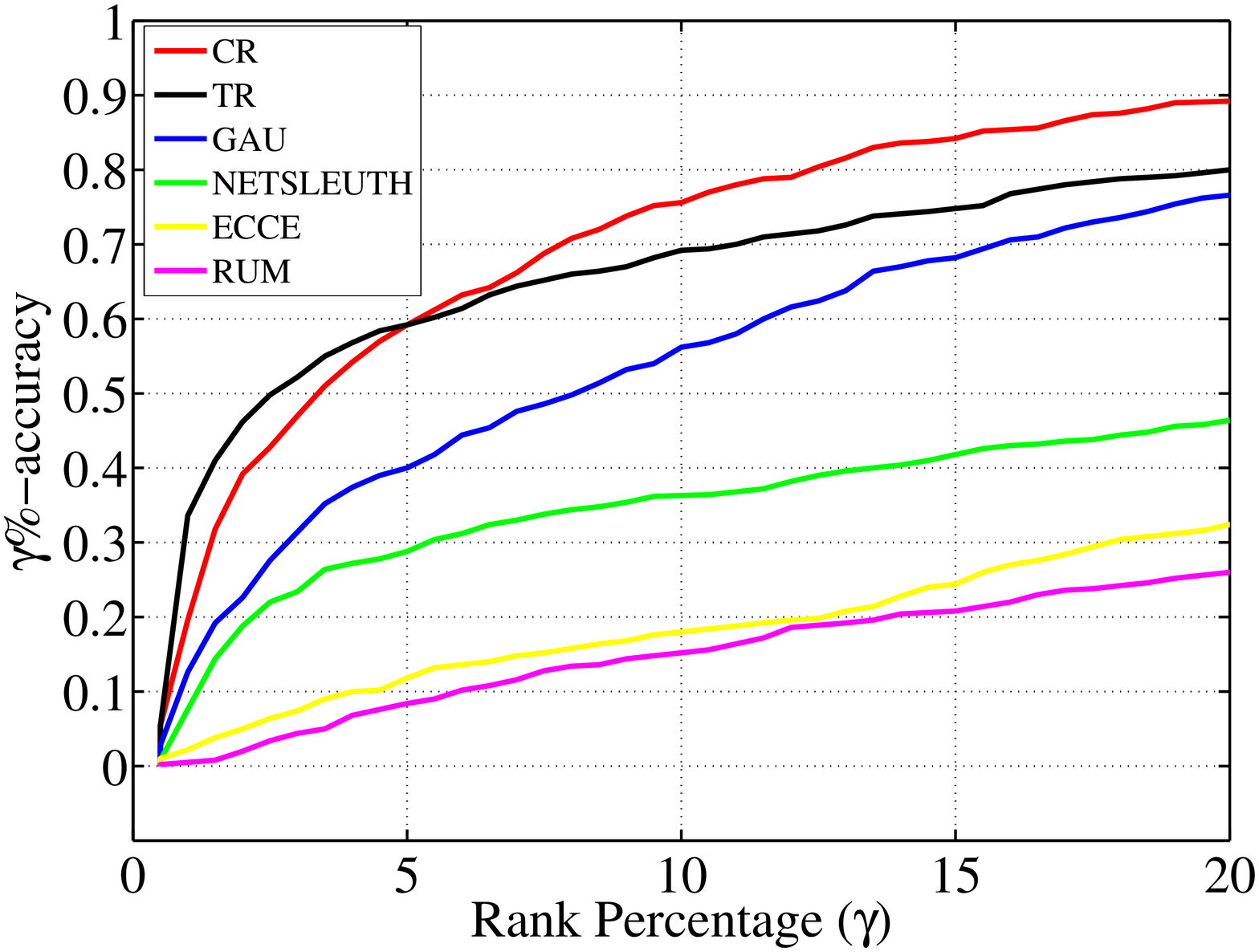}
                  \caption{The IAS network with $\mu=10$}\label{figure:CompCentralityIASmu10}
        \end{subfigure}
~
        \begin{subfigure}[b]{0.32\textwidth}
                \centering
                  \includegraphics[width=\textwidth]{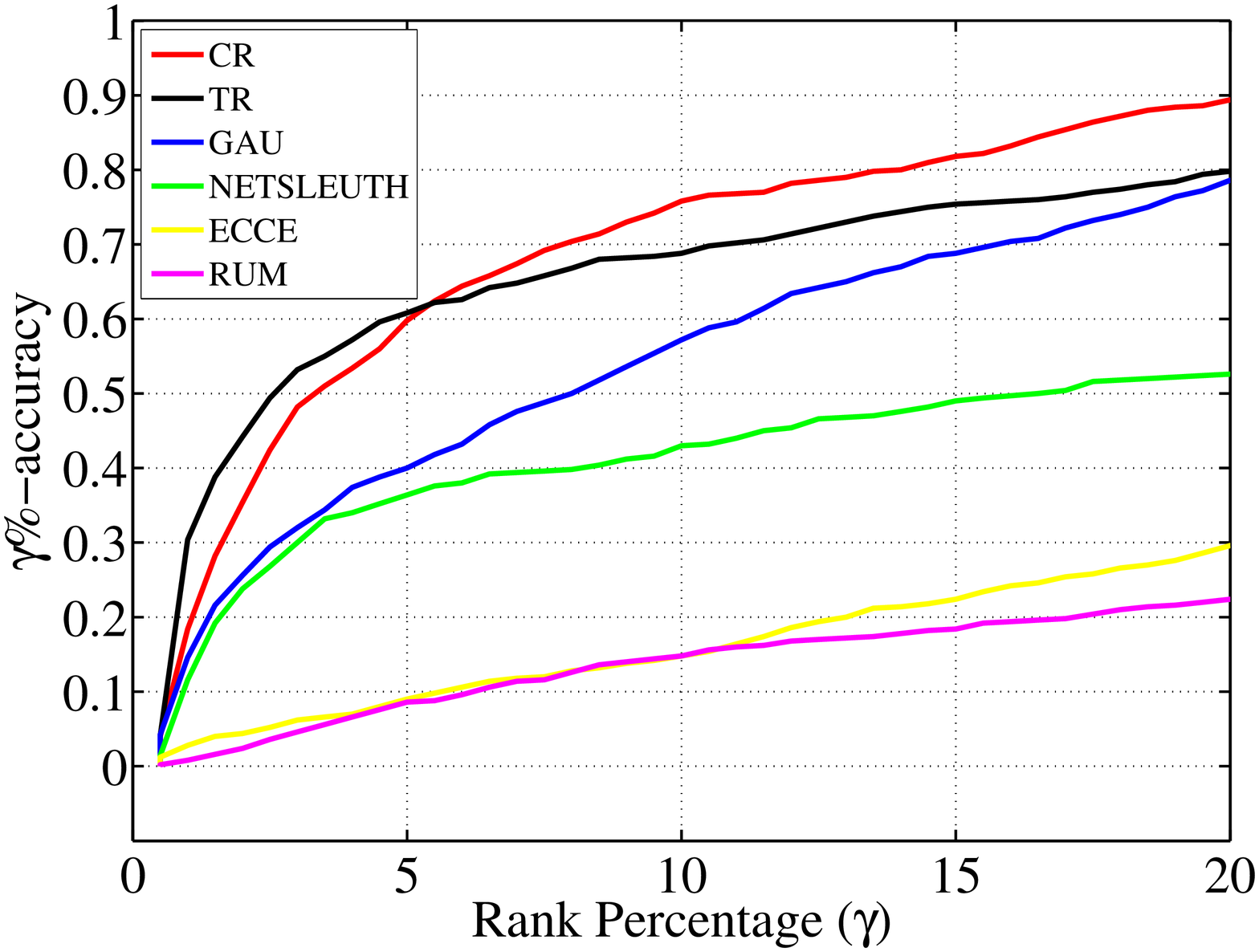}
                  \caption{The IAS network with $\mu=100$}\label{figure:CompCentralityIASmu100}
        \end{subfigure}
        \\
        \centering
        \begin{subfigure}[b]{0.32\textwidth}
                \centering
                \includegraphics[width=\textwidth]{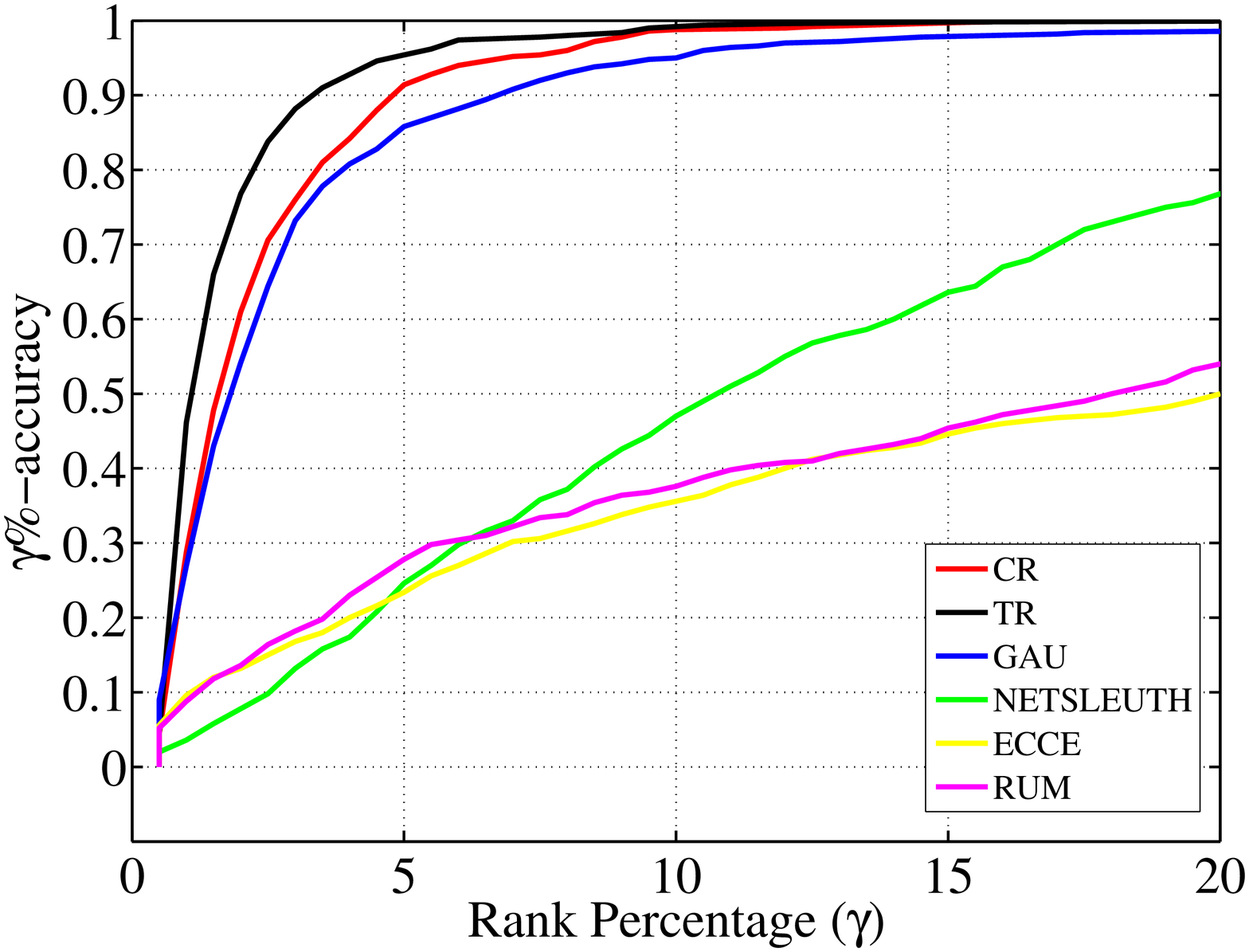}
                  \caption{The PG network with $\mu=1$}\label{figure:CompCentralityPGmu1}
        \end{subfigure}
~
        \begin{subfigure}[b]{0.32\textwidth}
                \centering
                  \includegraphics[width=\textwidth]{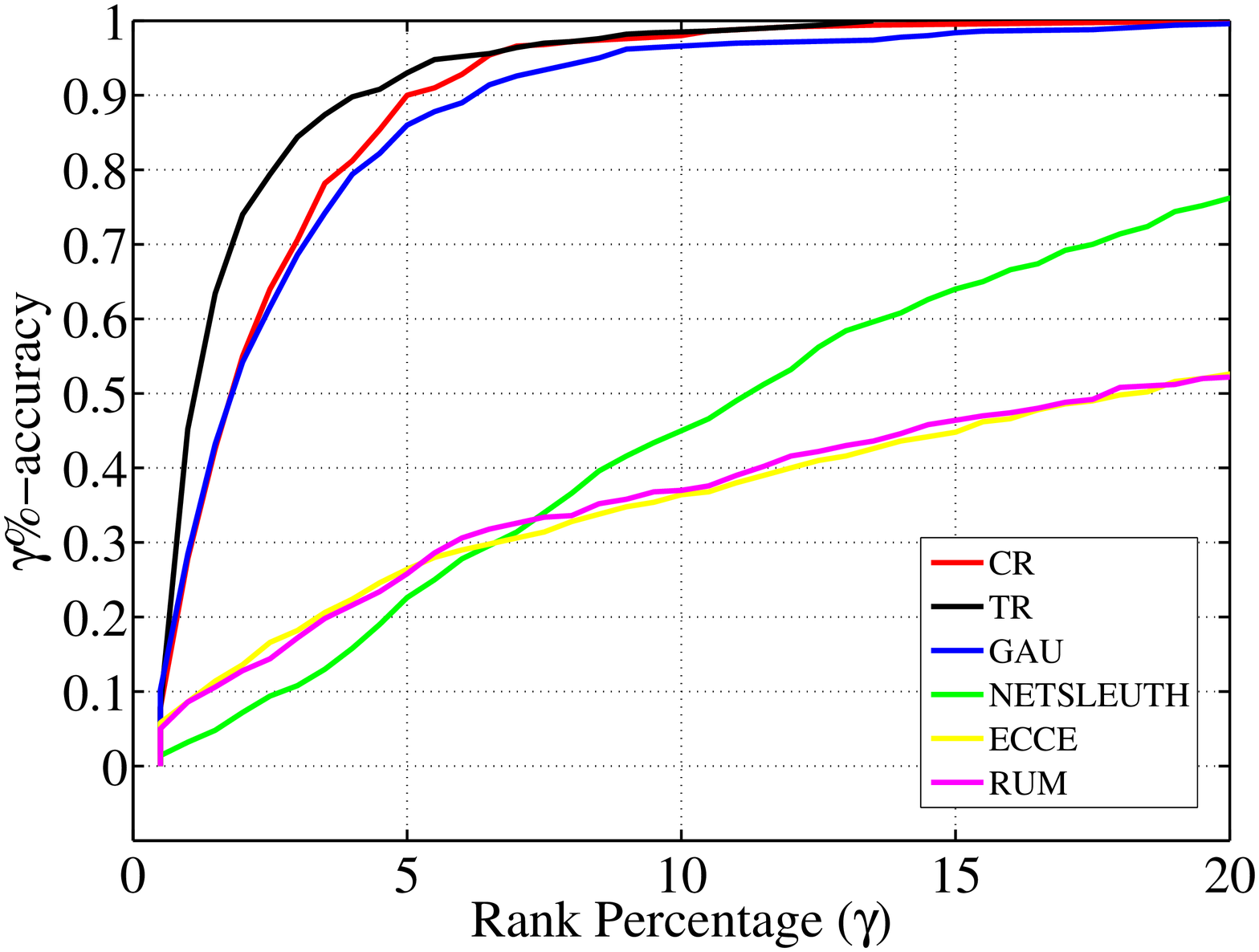}
                  \caption{The PG network with $\mu=10$}\label{figure:CompCentralityPGmu10}
        \end{subfigure}
~
        \begin{subfigure}[b]{0.32\textwidth}
                \centering
                  \includegraphics[width=\textwidth]{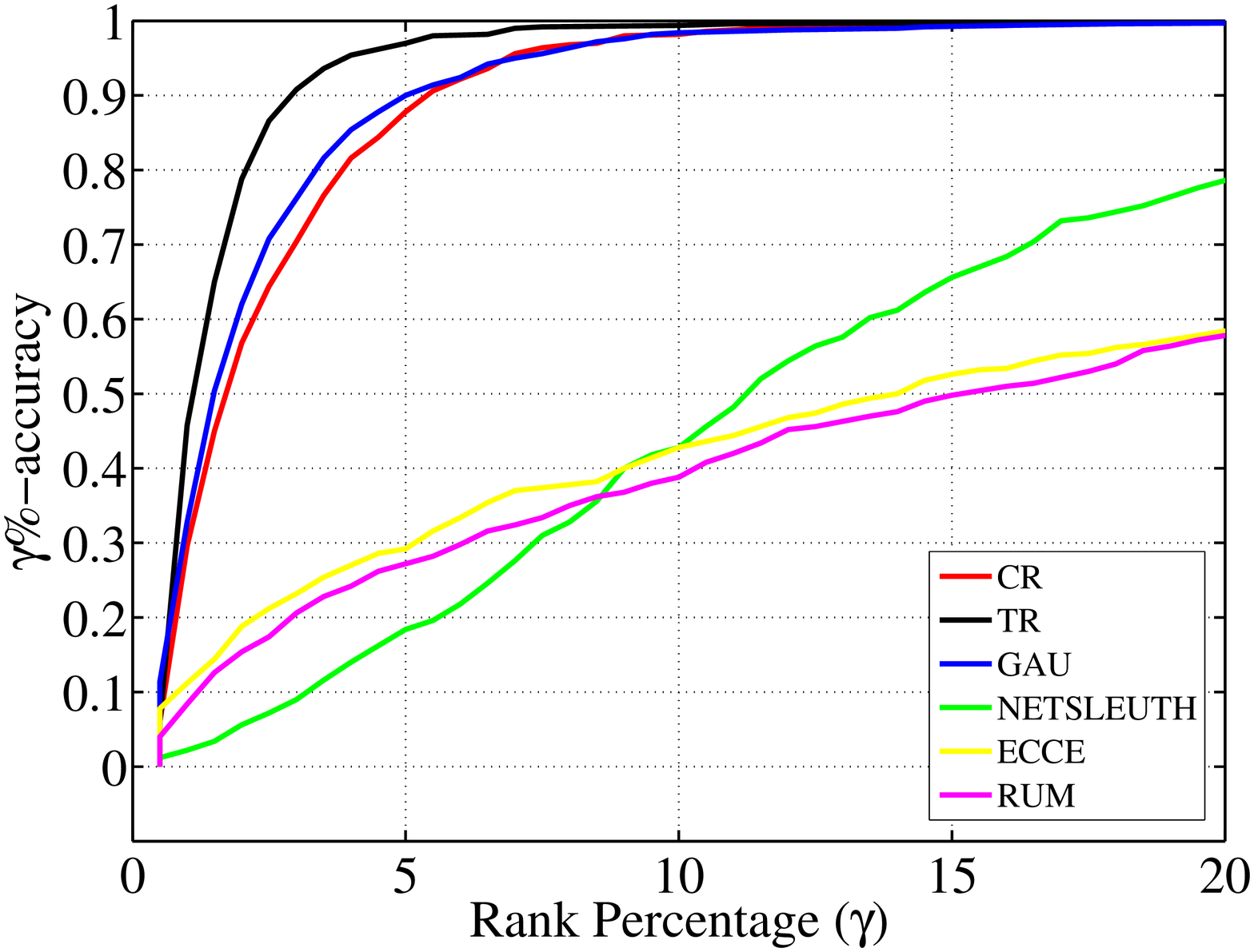}
                  \caption{The PG network with $\mu=100$}\label{figure:CompCentralityPGmu100}
        \end{subfigure}\\        
        \centering
        \caption{Comparison with Existing Algorithms with 50\% timestamps}\label{figure:differentmuCDF}
\end{figure*}

The results on the IAS and PG networks are presented in Figure \ref{figure:differentmuCDF} where the performance are consistent for different $\mu$ values. Recall that RUM, ECCE and NETLEUTH only use topological information.

\begin{itemize}

\item {\bf Observation 1:} In both networks, CR and TR perform much better than the other algorithms in the IAS network. In PG network, TR, CR and GAU have similar performance which dominates other algorithms due to the utilization of the timestamp information. In particular, in the IAS network, the $10\%$-accuracy of CR is 0.76 while $10\%$-accuracy of GAU and NETSLEUTH is 0.57 and 0.43, respectively when $\mu=100$.In the PG network, the $10\%$-accuracy of TR is 0.99 while that of GAU and NETSLEUTH is 0.98 and 0.43, respectively.

\item {\bf Observation 2:} Most algorithms, except NETSLEUTH, have higher $\gamma\%$-accuracy in the PG network than in the IAS network. We conjecture that it is because the IAS network has a small diameter and contains hub nodes while the PG network is more tree-like.

\item {\bf Observation 3:} NETSLEUTH dominates ECCE and RUM in the IAS network, but performs worse than ECCE and RUM in the PG network when $\gamma\leq 10.$ Furthermore, while all other algorithms have higher $\gamma$-accuracy in IAS than in PG, NETSLEUTH has lower $\gamma$-accuracy in IAS than in PG when $\gamma<10.$ A similar phenomenon will be observed in a later simulation as well.

\item {\bf Observation 4:} CR performs better in the IAS network when $\gamma\geq 5$ while TR performs better in the PG network.
\end{itemize}

\subsection{The Impact of Timestamp Distribution}\label{sec:ImpactOnTimestamp}

In the previous set of simulations, the revealed timestamps were uniformly chosen from all timestamps except the timestamp of the source, which was always excluded. We call this {\em unbiased distribution.} In this set of experiments, we study the impact of the distribution of the timestamps. We compared the unbiased distribution with a distribution under which nodes with larger infection time are selected with higher probability. In particular, we selected nodes iteratively. Let ${\cal N}^k$ denote the set of remaining infected nodes after selecting $k$ nodes, then the probability that node $i$ is selected in the next step is
\[
p_i^{(k)}= \frac{t_i-t_s}{\sum_{j\in {\cal N}^k}(t_j-t_s)},
\]
where $t_s$ is the infection time of the source. We call this  \emph{time biased distribution}.

\begin{figure*}[!t]
        \centering
        \begin{subfigure}[b]{0.3\textwidth}
                \centering
                \includegraphics[width=\textwidth]{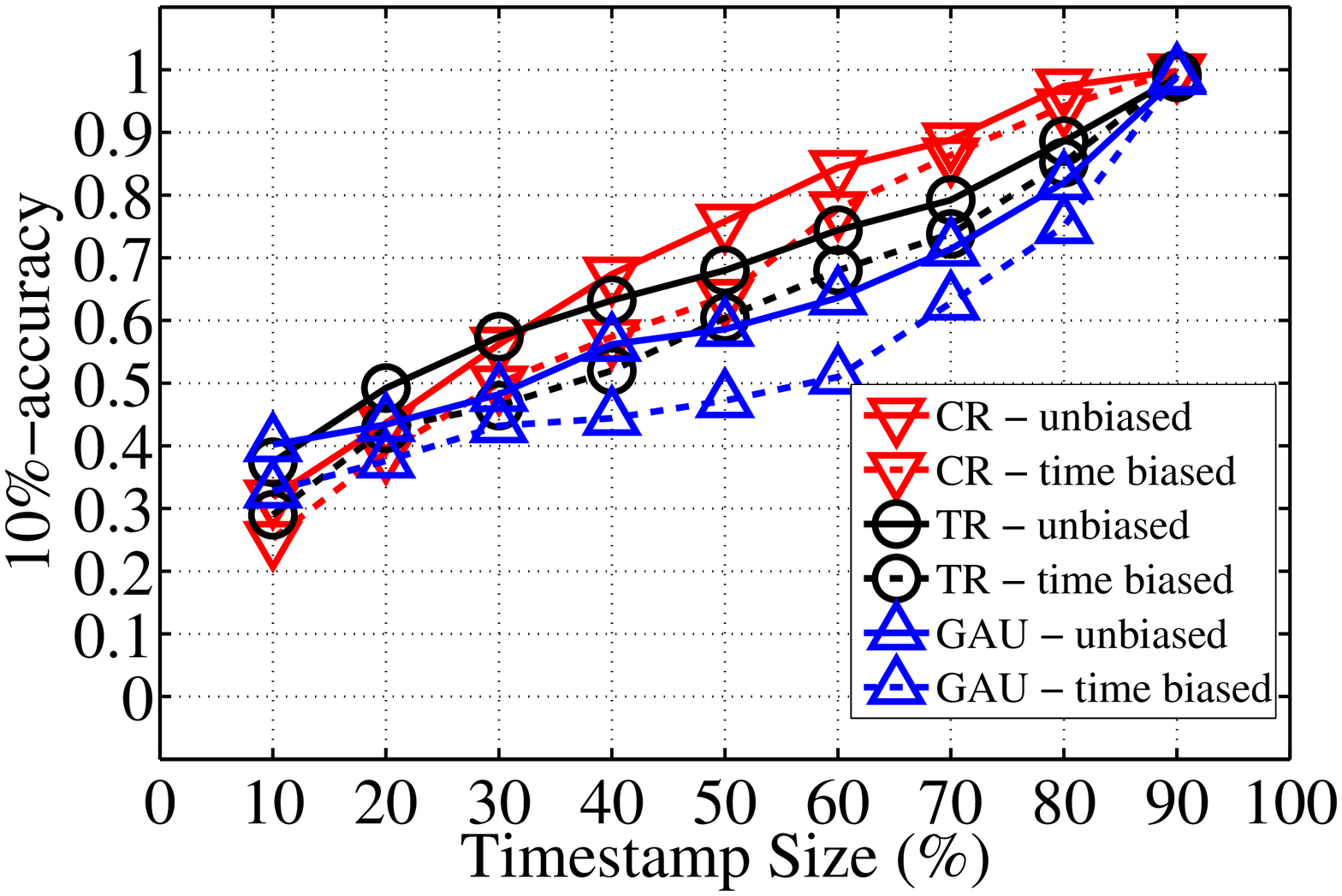}
                  \caption{The IAS network with $\mu=1$}\label{figure:IASsamplemethodmu1}
        \end{subfigure}
~
    \begin{subfigure}[b]{0.3\textwidth}
                \centering
                \includegraphics[width=\textwidth]{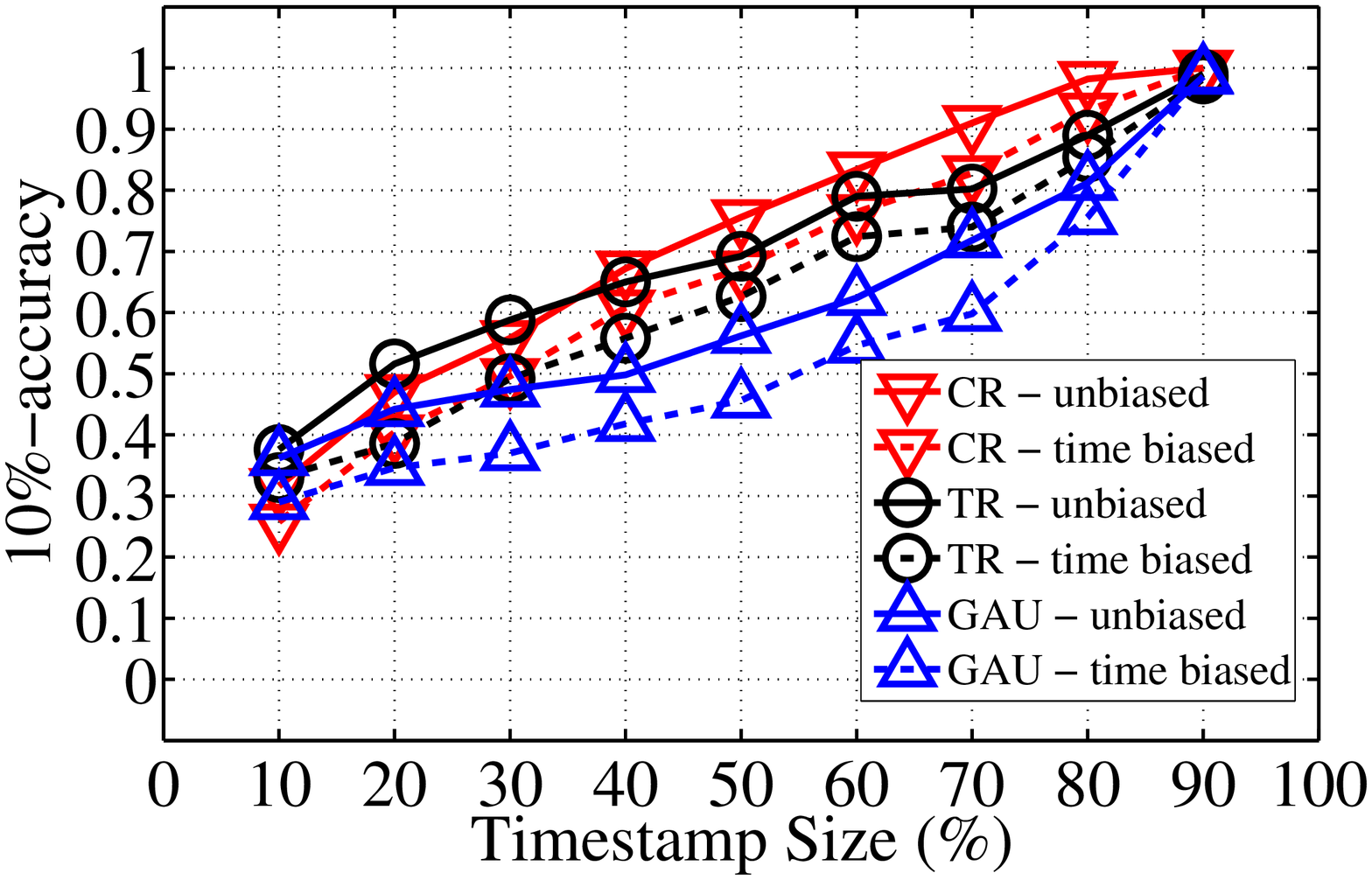}
                  \caption{The IAS network with $\mu=10$}\label{figure:IASsamplemethodmu10}
        \end{subfigure}
~
    \begin{subfigure}[b]{0.3\textwidth}
                \centering
                \includegraphics[width=\textwidth]{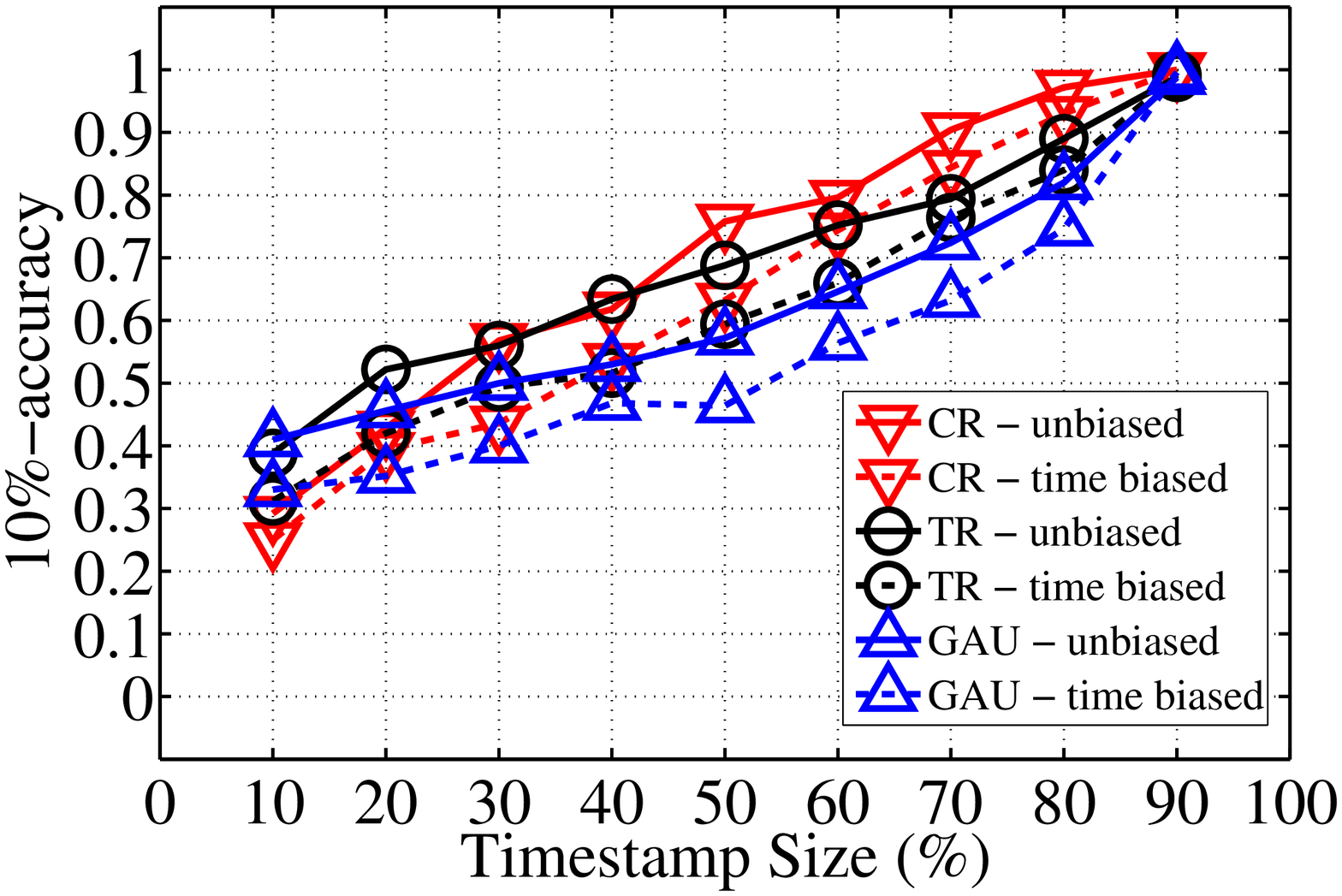}
                  \caption{The IAS network with $\mu=100$}\label{figure:IASsamplemethodmu100}
        \end{subfigure}
       \\
        \begin{subfigure}[b]{0.3\textwidth}
                \centering
                  \includegraphics[width=\textwidth]{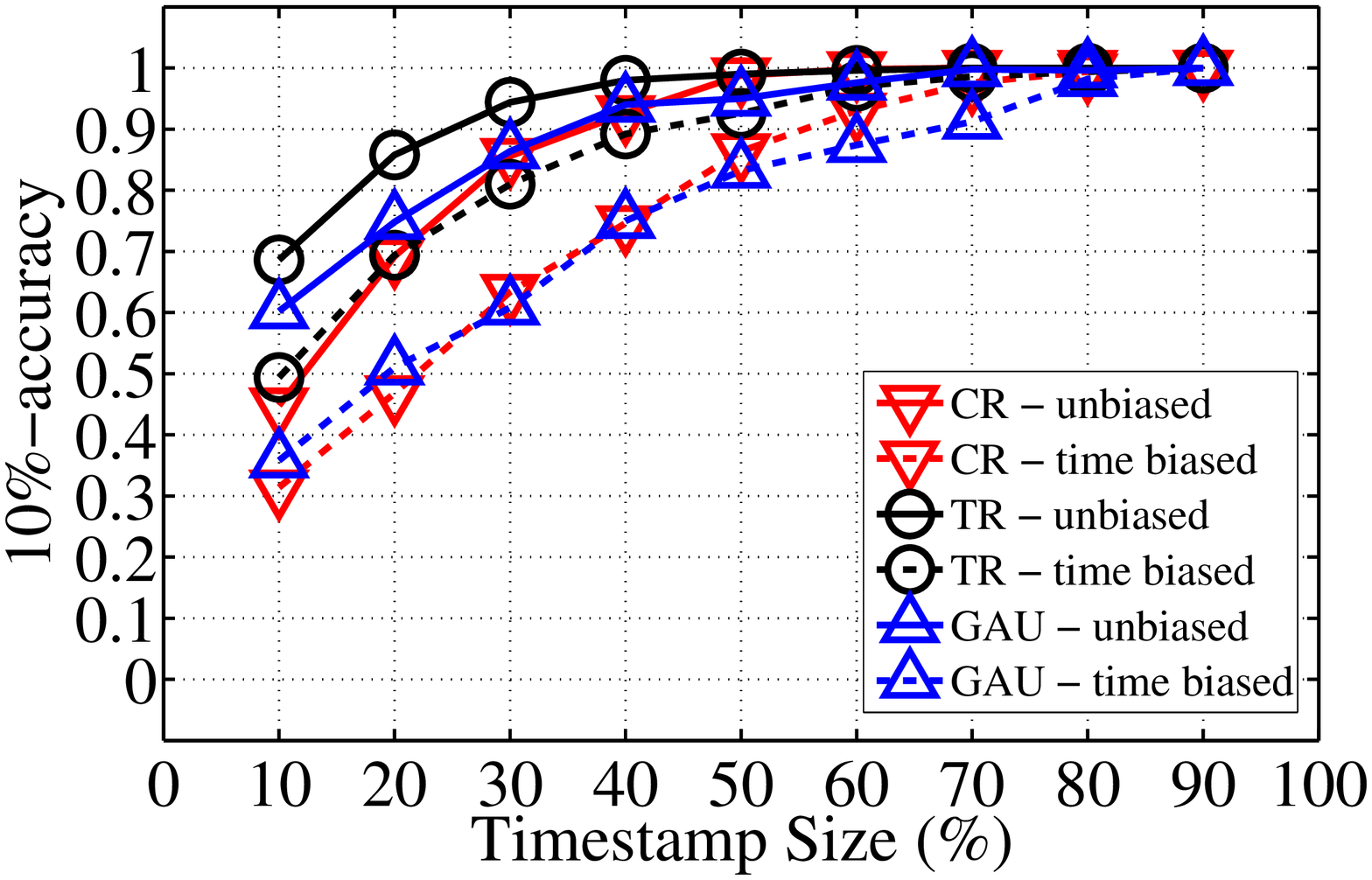}
                  \caption{The PG network with $\mu=1$}\label{figure:PGsamplemethodmu1}
        \end{subfigure}
~
        \begin{subfigure}[b]{0.3\textwidth}
                \centering
                  \includegraphics[width=\textwidth]{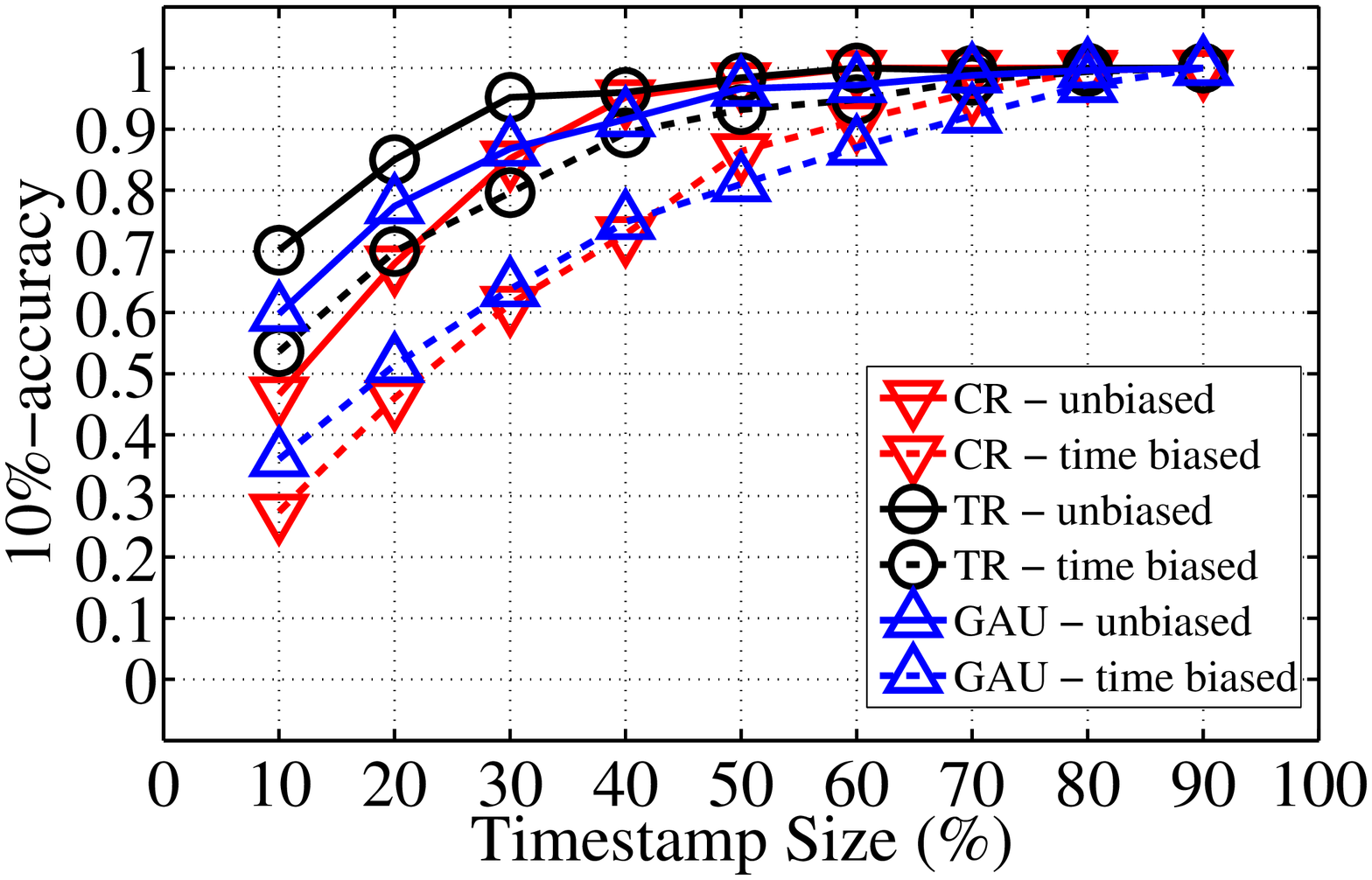}
                  \caption{The PG network with $\mu=10$}\label{figure:PGsamplemethodmu10}
        \end{subfigure}
~       \begin{subfigure}[b]{0.3\textwidth}
                \centering
                  \includegraphics[width=\textwidth]{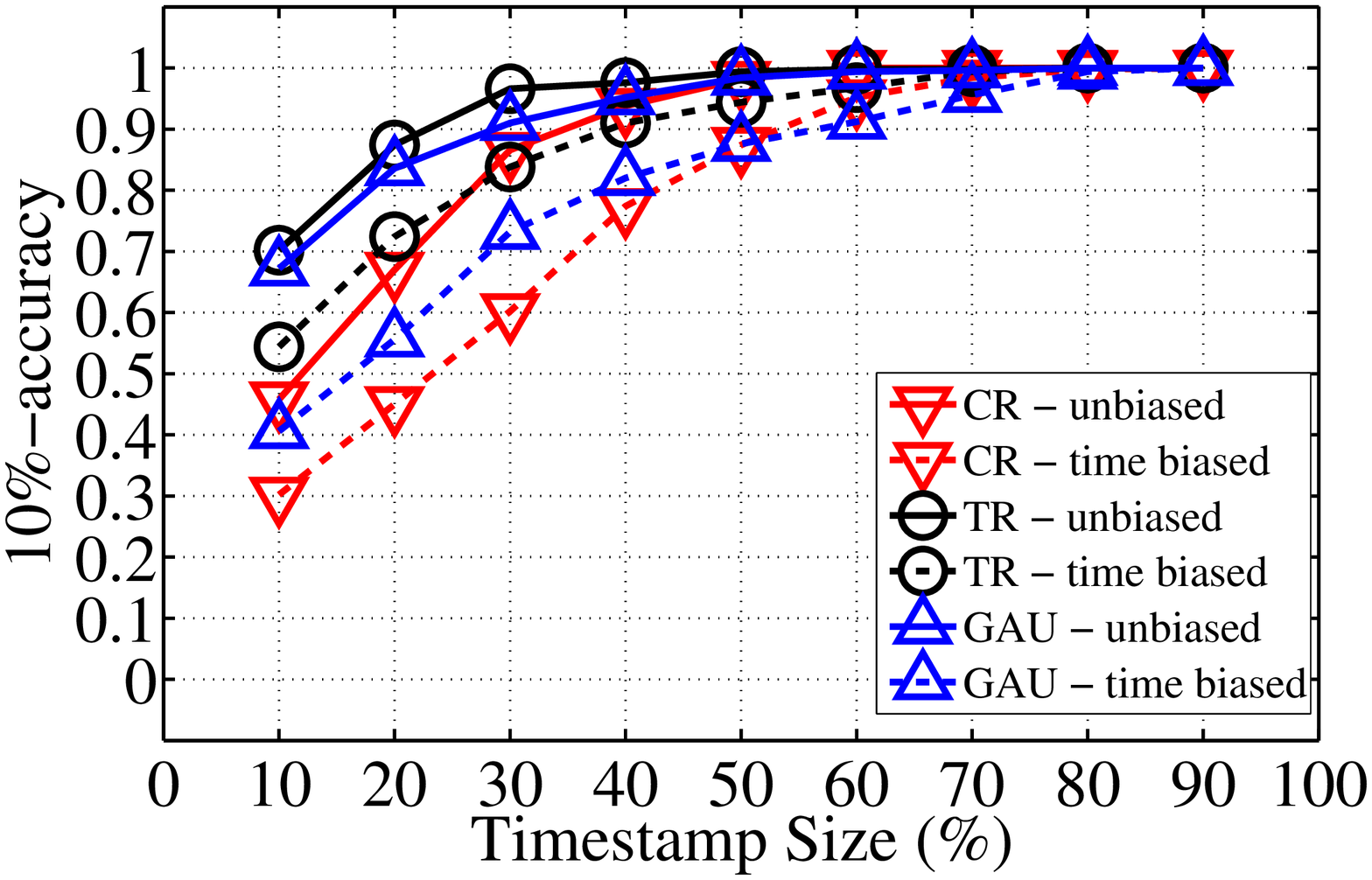}
                  \caption{The PG network with $\mu=100$}\label{figure:PGsamplemethodmu100}
        \end{subfigure}
        \centering
        \caption{The Impacts of the Distribution and Size of Timestamps}\label{figure:DifferentMu}
\end{figure*}

In this section, we evaluated the performance of our algorithms and GAU with different sizes of observed timestamps and different distributions of the observed timestamps. All the experiment setups are the same as in Section \ref{sec:comGraphCentrality}. We evaluate the algorithms with $\mu=\{1,10,100\}$ and the results of different number of timestamps are shown in Figure \ref{figure:DifferentMu}.


 Note that the performance of RUM, ECCE and NETSLEUTH are independent of timestamp distribution and size, so we did not include these algorithms in the figures. From the figure, we have the following observations:
\begin{itemize}
\item {\bf Observation 5:} We varied the size of observed timestamps from 10\% to 90\%. As we expected, the $\gamma\%$-accuracy increases as the size increases under both CR and TR. Interestingly, in the IAS network, the $10\%$-accuracy of GAU is worse than TR and CR when more than $20\%$ of the timestamps are observed. We conjecture this is because in small world networks such as the IAS network, the spreading tree is very different from the breadth-first search tree rooted at the source. Since GAU always uses the breadth-first search trees regardless of the size of timestamps, more timestamps do not result in a more accurate spreading tree.  The spreading tree constructed by EIF, on the other hand, depends on the size of timestamps and is more accurate as the size of timestamps increases.


\item {\bf Observation 6:} In both networks, the time-biased distribution results in $5\%$ to $15\%$ reduction of the $\gamma\%$-accuracy. This shows that earlier timestamps provide more valuable information for locating the source. However, the trends and relative performance of the three algorithms are similar to those in the unbiased case.

\item {\bf Observation 7:} CR performs better in the IAS network when the timestamp size is larger than 40\%; and TR performs better in the PG network.

\item {\bf Observation 8:} The $\gamma\%$-accuracy is much higher in the PG network than that in the IAS network under both the unbiased distribution and time-biased distribution. For example, with the time-biased distribution and  20\% of timestamps, the $10\%$-accuracy of TR is 0.87 in PG and is only 0.52 in IAS when $\mu=100$. This again confirms that the source localization problem is more difficult in networks with small diameters and hub nodes.

\end{itemize}

\subsection{The Impact of the Diffusion Model}
\label{sec:spikem}

\begin{figure*}[!t]
\centering
        \begin{subfigure}[b]{0.4\textwidth}
                \centering
                \includegraphics[width=\textwidth]{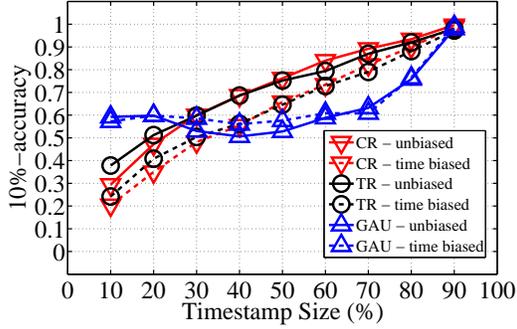}
                  \caption{The IAS Network under the IC Model}\label{figure:IASIC}
        \end{subfigure}
~    \begin{subfigure}[b]{0.4\textwidth}
                \centering
                \includegraphics[width=\textwidth]{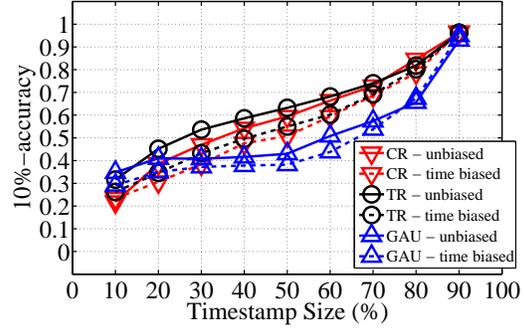}
                  \caption{The IAS Network under the SpikeM Model}\label{figure:IASsparkm}
        \end{subfigure}
       \\
        \centering
    
         \begin{subfigure}[b]{0.4\textwidth}
                        \centering
                          \includegraphics[width=\textwidth]{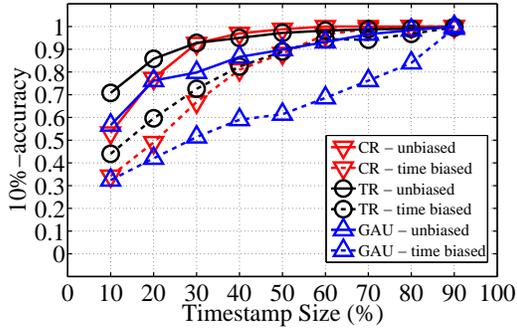}
                          \caption{The PG Network under the IC Model}\label{figure:PGIC}
                \end{subfigure}
~
        \begin{subfigure}[b]{0.4\textwidth}
                \centering
                  \includegraphics[width=\textwidth]{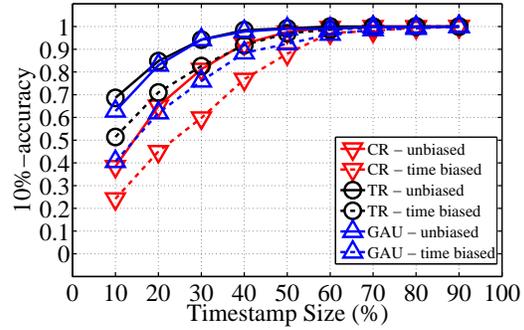}
                  \caption{The PG Network under the SpikeM Model}\label{figure:PGsparkm}
        \end{subfigure}
        \centering
        \caption{The Performance of CR, TR and GAU under different diffusion models}\label{figure:differentdiffusion}
\end{figure*}

In all previous experiments, we used the truncated Gaussian model for contagion. We now study the robustness of CR and TR to the contagion models. We conducted the experiments using the IC model \cite{KemKleTar_03} and SpikeM model \cite{MatSakPra_12} for contagion. \textcolor{black}{Both models are time slotted, so are very different from the truncated Gaussian model. In the IC model, each infected node has only one chance to infect each of its neighbors. If the infection failed, the node cannot make more attempts. In the experiments, the infection probability along each edge is selected with a uniform distribution over $(0,1).$} SpikeM model has been shown to match the patterns of real-world information diffusion well. {\color{black} In the SpikeM model, infected nodes become less infectious as the increase of the time. Furthermore, the activity level of a user in different time periods of a day varies to match the rise and fall patterns of information diffusion in the real world.}  In our experiments, we used the parameter set C5 in Table 3 of \cite{MatSakPra_12} which was obtained based on MemeTracker dataset. The results are shown in Figure \ref{figure:differentdiffusion}, where in each figure, the size of timestamps varies from 10\% to 90\%.

\begin{itemize} 
\item {\bf Observation 9:} \textcolor{black}{Under both the IC and SpikeM models, the GAU algorithm has a better performance when less than $20\%$ timestamps are observed in the IAS network. The performance of TR and CR dominate GAU when more than $20\%$ timestamps are observed. For the PG network, the performances of TR and CR are better than GAU under th IC model, and the performance of TR is better than GAU under the SpikeM model.}
\end{itemize}
\textcolor{black}{{\bf Remark 5:} Another popular diffusion model is the Linear Threshold (LT) model\cite{KemKleTar_03}. However, in the experiments, we found that it is difficult for a single source to infect more than $150$ nodes under the LT model. Therefore, we only conducted experiments with the IC model.}

\subsection{The Impact of Network Topology}
\begin{figure}[htb]
\begin{centering}
  \includegraphics[width=0.5\textwidth]{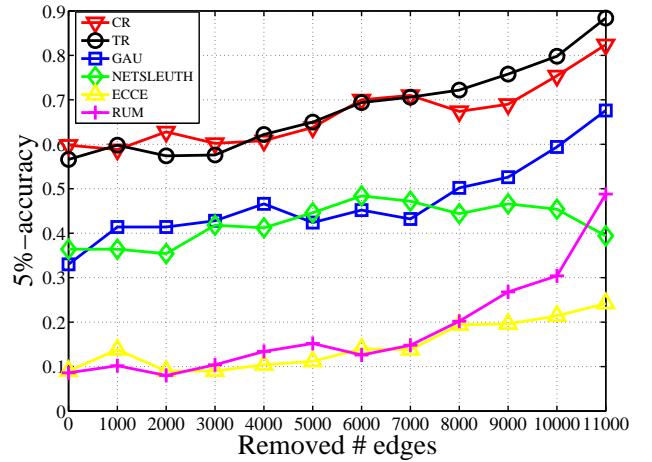}
  \caption{The $\gamma\%$-Accuracy as the Number of Removed Edges Increases}\label{figure:DeleteEdgeCDF}
 \end{centering}
\end{figure}

In the previous simulations, we have observed that locating the source in the PG network is easier than in the IAS network.  We conjecture that it is because the IAS network is a small-world network while the PG network is more tree-like.  To verify this conjecture,  we removed edges from the IAS network to observe the change of $\gamma\%$-accuracy as the number of removed edges increases. For each removed edge, we randomly picked one edge and removed it if the network remains to be connected after the edge is removed.  We used the truncated Gaussian model and all other settings are the same as those in Section \ref{sec:comGraphCentrality}. The results are shown in Figure \ref{figure:DeleteEdgeCDF}.
\begin{itemize}
\item {\bf Observation 10:} After removing 11,000 edges, the ratio of the number of edges to the number of nodes is $11,002/10,670=1.03,$ so the network is tree-like. As showed in Figure \ref{figure:DeleteEdgeCDF}, the 5\%-accuracy of all algorithms, except NETSLEUTH,  improves as the number of the removed edges increases, which confirms our conjecture. The 5\%-accuracy of NETSLEUTH starts to decrease when the number of removed edges is more than $6,000.$ This is consistent with the observation we had in Figure \ref{figure:differentmuCDF}, in which the 5\% accuracy of NETSLUETH in PG is worse than that in IAS.
\end{itemize}

\subsection{Weibo Data Evaluation}
In this section, we evaluated the performance of our algorithms with real-world network and real-world information spreading. The dataset is the Sina Weibo\footnote{\url{http://www.weibo.com/}} data, provided by the WISE 2012 challenge\footnote{\url{http://www.wise2012.cs.ucy.ac.cy/challenge.html}}. Sina Weibo is the Chinese version of Twitter, and the dataset includes a friendship graph and a set of tweets. 

\begin{figure*}[!t]
        \centering
        \begin{subfigure}[b]{0.45\textwidth}
                \centering
                \includegraphics[width=\textwidth]{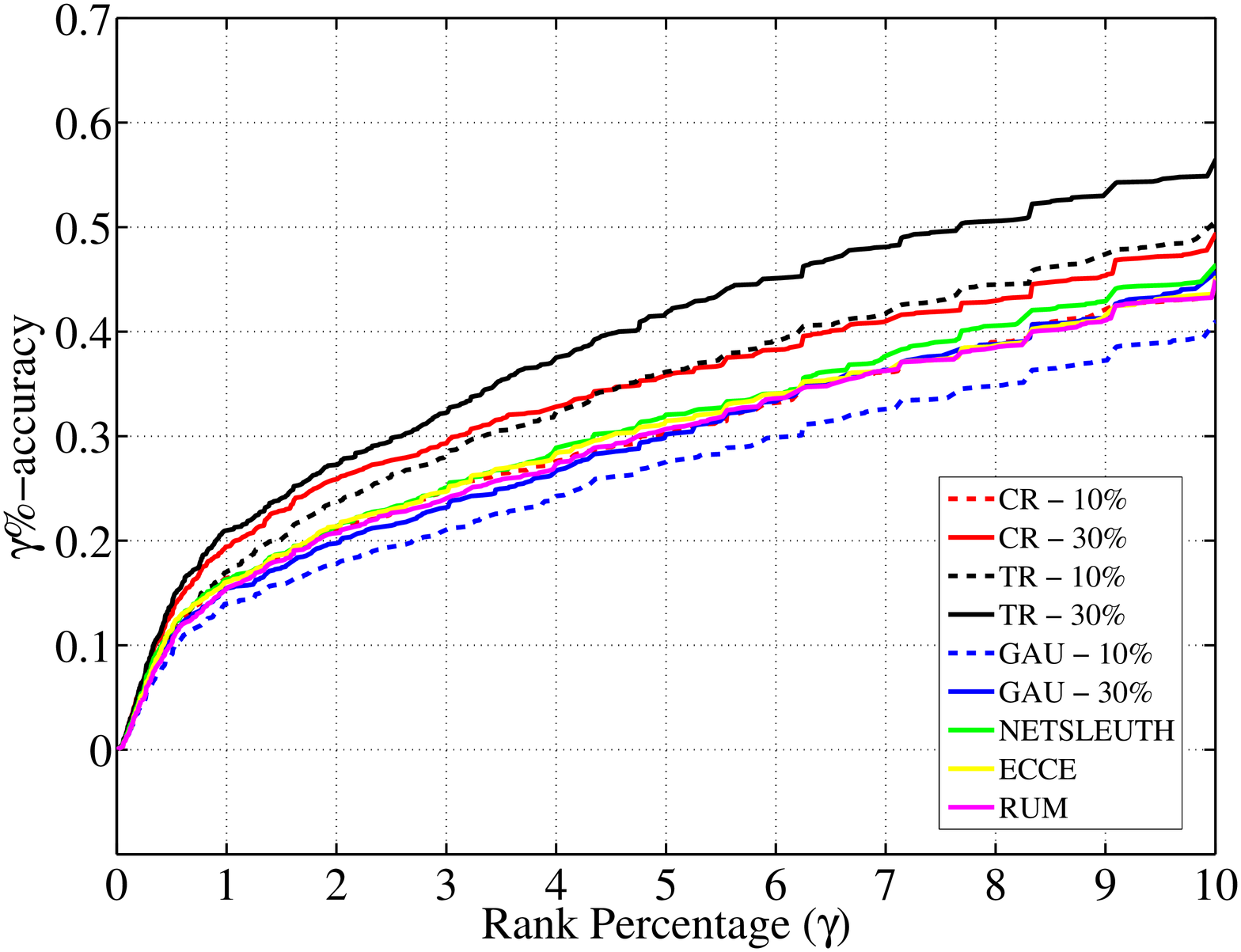}
                  \caption{All Tweets}\label{figure:WeiboPerformanceAll}
        \end{subfigure}
~
        \begin{subfigure}[b]{0.45\textwidth}
                \centering
                  \includegraphics[width=\textwidth]{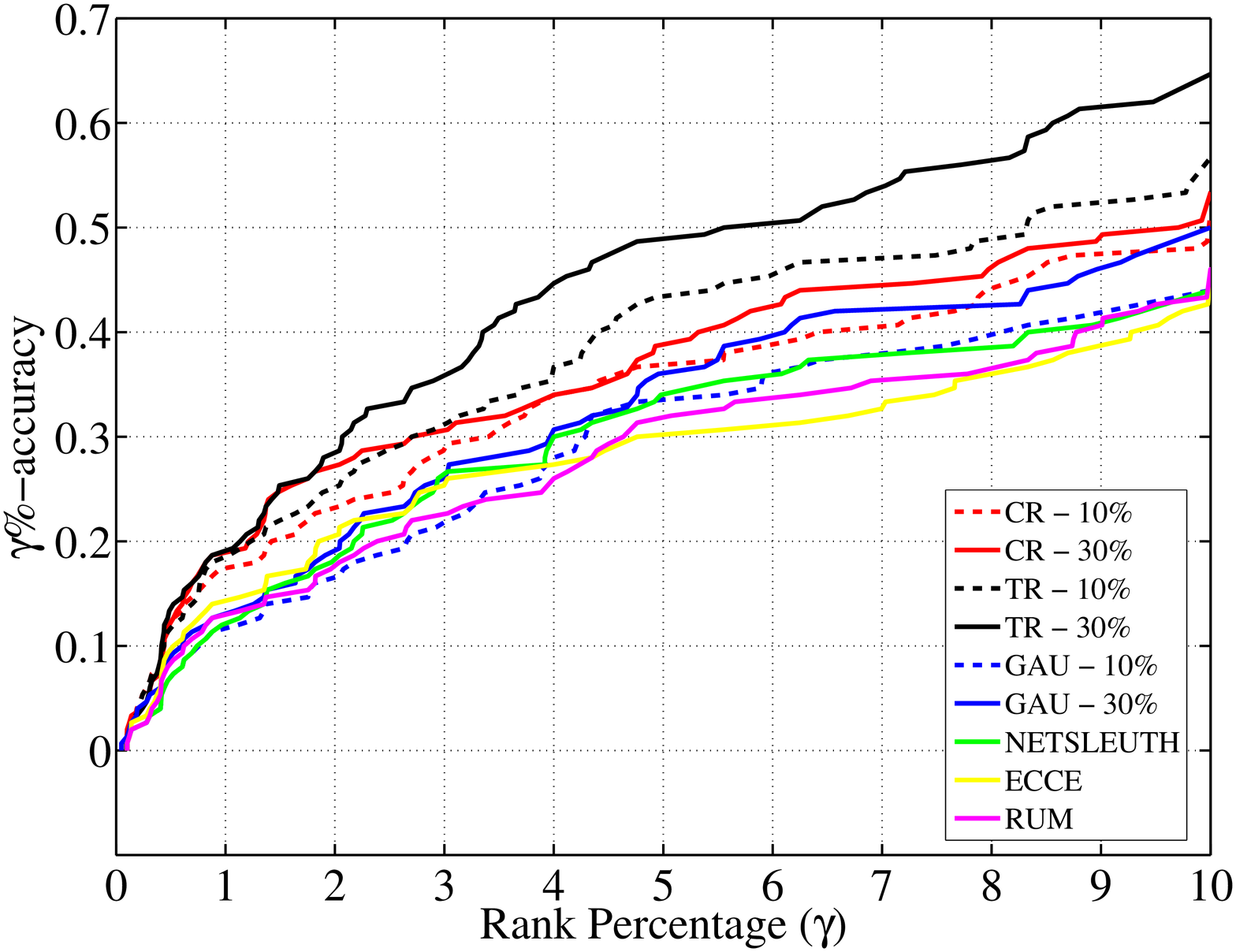}
                  \caption{Resample by Degree}\label{figure:WeiboPerformanceDegree}
        \end{subfigure}
        \centering
        \caption{Performance on Weibo Data}\label{figure:WeiboEvalTotal}
\end{figure*}
The friendship graph is a directed graph with 265,580,802 edges and 58,655,849 nodes. The tweet dataset includes 369,797,719 tweets. Each tweet includes the user ID and post time of the tweet. If the tweet is a retweet of some tweet, it includes the tweet ID of the original tweet, the user who post the original tweet, the post time of the original tweet, and the retweet path of the tweet which is a sequence of user IDs. For example, the retweet path $a\rightarrow b\rightarrow c$ means that user $b$ retweeted user $a$'s tweet, and user $c$ retweeted user $b$'s.

We selected the tweets with more than 1,500 retweets. For each tweet,  all users who retweet the tweet are viewed as infected nodes and we extracted the subnetwork induced by these users. We also added those edges on the retweet paths to the subnetwork if they are not present in the friendship graph, by treating them as missing edges in the friendship network. The user who posts the original tweet is regarded as the source. If there does not exist a path from the source to an infected node along which the post time is increasing, the node was removed from the subnetwork. In addition, to make sure we have enough timestamps, we remove the samples with less than 30\% timestamps.

After the above preprocessing, we have 1,170 tweets with at least 30\% observed timestamps. Similar to section  \ref{sec:comGraphCentrality} in the paper, we grouped the tweets into five bins according the degree of the source in the friendship graph. In the $k^{\rm th}$ bin (for $k=1, 2, 3, 4$),  the degree of the source is between $8000(k-1)$ to $8000k-1.$ In the $5^{\rm th}$ bin, the degree of the source is at least $32,000.$ The number of tweets in the bins are
$[ 568  \quad 147 \quad   70 \quad   68  \quad  317 ].$ From each bin, we draw 30 samples without replacement. For completeness, we also evaluated the performance with all 1,170 tweets. The results are summarized in Figure \ref{figure:WeiboEvalTotal}. Figure \ref{figure:WeiboPerformanceAll} shows the performance with all tweets samples and Figure \ref{figure:WeiboPerformanceDegree} shows the performance if we resample the tweets by the above degree bins. The observed timestamps are uniformly selected from the available timestamps and the source node is excluded.

\begin{itemize}

\item {\bf Observation 11:} Figure \ref{figure:WeiboEvalTotal} shows that CR and TR  dominates GAU with both $10\%$ and $30\%$ of timestamps. In particular for the resample by degree case, TR performs very well and dominates all other algorithms with a large margin. The $10\%$-accuracy of TR with 30\% timestamps is around 0.64 while that of CR is 0.53 and that of NETSLEUTH is only 0.4.
\end{itemize}

{\bf Summary:} From the synthetic data and real data evaluations, we have seen that both TR and CR perform better than existing algorithms, and are robust to diffusion models and timestamp distributions. Furthermore, TR performs better than CR in most cases. CR performs better than TR only in the IAS network when the sample size is large ($\geq 30\%$ under the truncated Gaussian diffusion, $\geq 50\%$ under the IC model and $\geq 70\%$ under the SpikeM model). Therefore, {\em we would recommend TR for general cases.} 

%% file: report/Conclusion.tex
\section{Conclusions and Extensions}
\label{sec:con}
In this paper, we studied the problem of locating the contagion source with partial timestamps. We developed two ranking algorithms, CR and TR. Experimental evaluations on synthetic and real-world data demonstrated that CR and TR improve the ranking accuracy significantly compared with existing algorithms, and perform well in real-world networks.
\begin{figure*}
\begin{minipage}{0.48\linewidth}
\begin{centering}
  \includegraphics[width=3.8in]{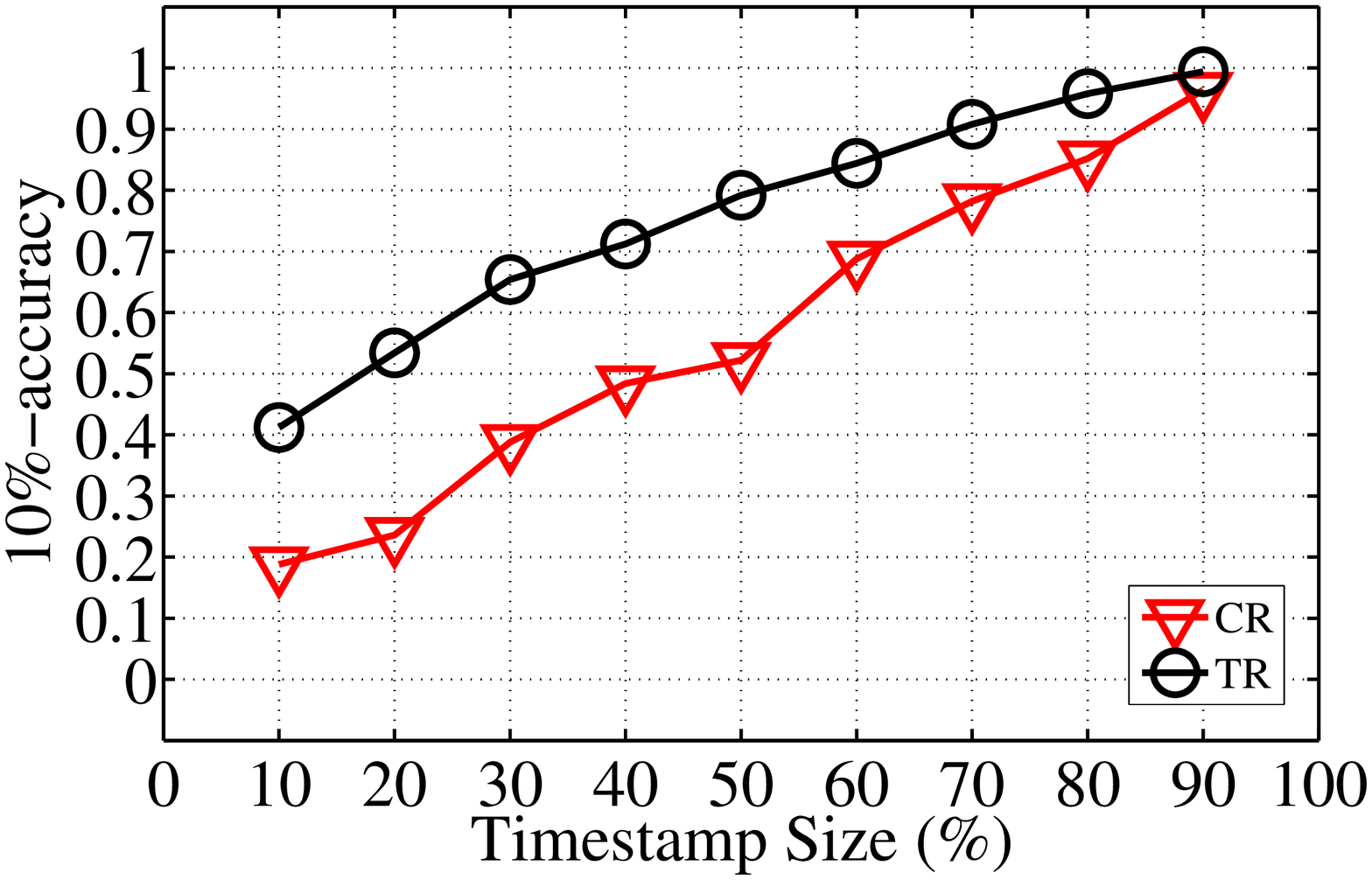}
  \caption{The Performance of CR, TR in the IAS Network under the SpikeM Model with Partially Observed Infected Nodes}\label{figure:ias80partial}
  \end{centering}
  \end{minipage}
\hfill
\begin{minipage}{0.48\linewidth}
\begin{centering}
  \includegraphics[width=3.8in]{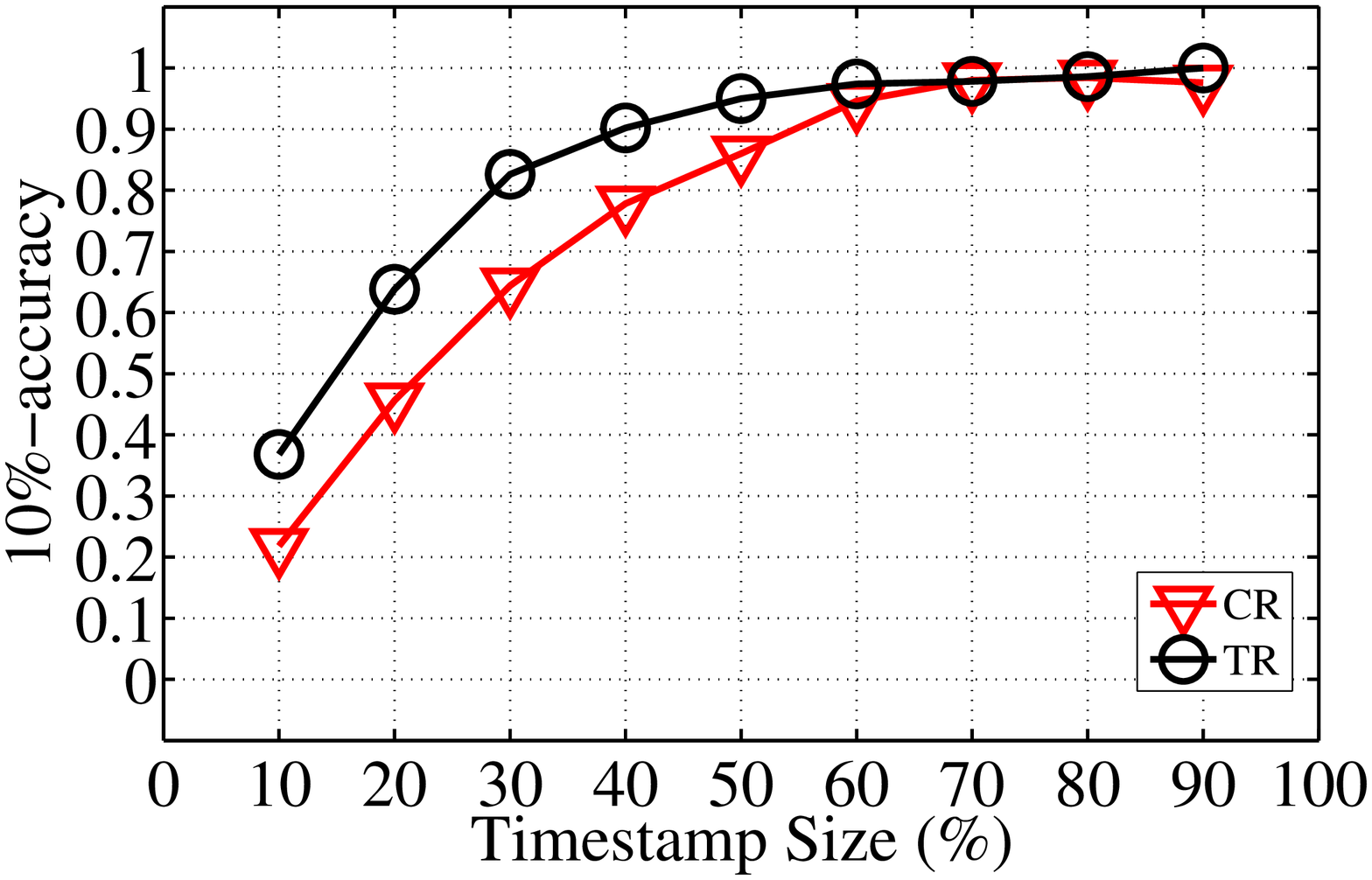}
  \caption{The Performance of CR, TR in the PG Network under the SpikeM Model with Partially Observed Infected Nodes}\label{figure:pg80partial}
  \end{centering}
  \end{minipage}
\end{figure*}
\subsection{Partially Observed Infected Nodes}
In Section \ref{sec:approach}, we assumed that $\cal I$ includes all infected nodes. This, however, is not a critical assumption. When $\cal I$ only includes a subset of infected nodes, call them observed infected nodes, CR and TR can be used to rank these nodes according to their likelihood of being the {\em earliest observed infected node.} Using the setting in Section \ref{sec:spikem} (the SpikeM model) and assuming 80\% of infected nodes are observed, figures \ref{figure:ias80partial} and \ref{figure:pg80partial} show the $\gamma\%$-accuracy of CR and TR  for locating the earliest observed infection nodes in the IAS and PG networks under the unbiased timestamps distribution. We can see that the performance is similar to Figures \ref{figure:IASsparkm} and \ref{figure:PGsparkm}. This demonstrates that our algorithm could be applied to the scheme where partially infected nodes are observed.

\subsection{Other Side Information}
\begin{figure}
\begin{minipage}{1\linewidth}
\begin{centering}
  \includegraphics[width=1.6in]{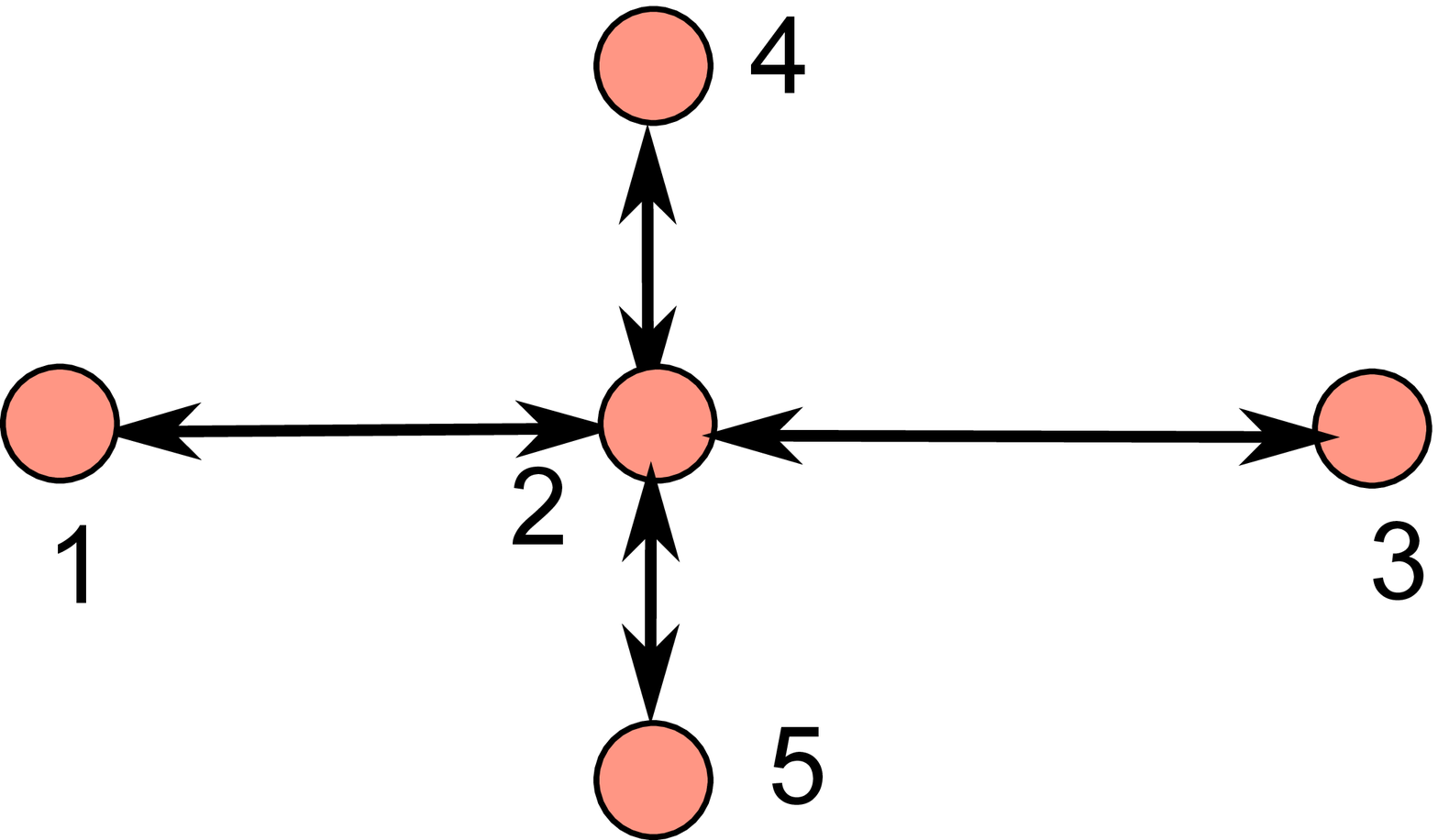}
  \caption{The Subnetwork before Modification}\label{figure:side}
  \end{centering}
\end{minipage}
\vfill
\begin{minipage}{1\linewidth}
\begin{centering}
  \includegraphics[width=1.6in]{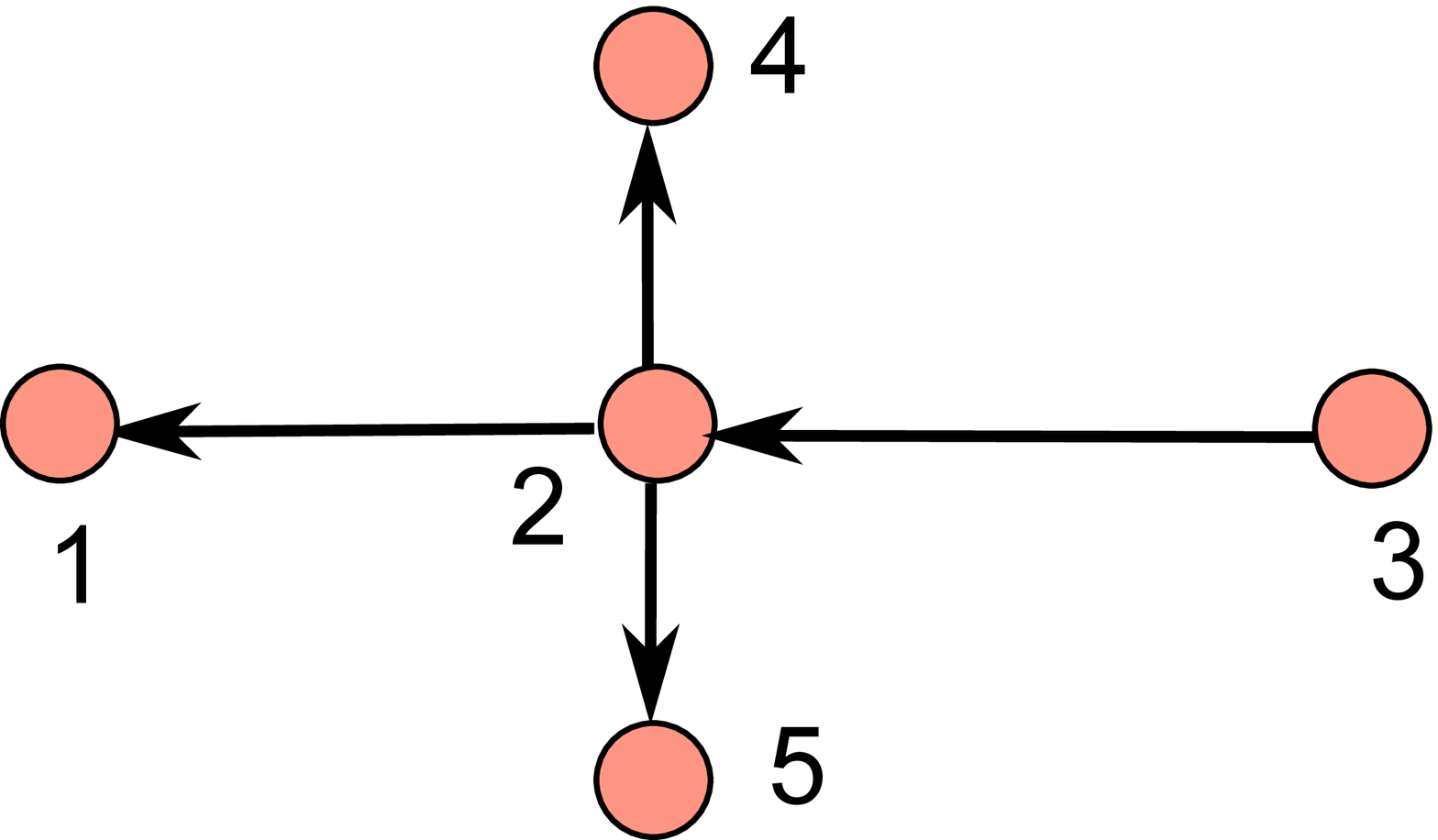}
  \caption{The Subnetwork after Including Information $6\rightarrow 7$}\label{figure:side2}
  \end{centering}
\end{minipage}
\end{figure}

In some practical scenarios, we may have other side information than timestamps such as {\em who infected whom.} This side information can be incorporated in the algorithm by modifying the network $G.$  Consider the example in Figure \ref{figure:side}. If we know that node $2$ was infected by node $3,$ then we can removed all incoming edges to node $2,$ except $3\rightarrow 2,$ and the edge $2\rightarrow 3$ to obtain a modified $G$ as shown in Figure \ref{figure:side2}. We can then apply CR and TR on the modified graph to rank the observed infected nodes.

%% file: report/JustificationoftheQuadraticCostFunction.tex
\section*{Proof of Theorem \ref{thm:np}}
\label{sec: NP hardness}

Assume all nodes in the network are infected nodes and the infection time of two nodes (say node $v$ and node $w$) are observed. Without loss of generality, assume $\InfObsTimeEnt_v<\InfObsTimeEnt_w.$ Furthermore, assume the graph is undirected (i.e., all edges are bidirectional) and $$|\InfObsTimeEnt_v-\InfObsTimeEnt_w|\geq \mu |\InfNode|.$$ We will prove the theorem by showing that computing the cost of node $v$ is related to the longest path problem between nodes $v$ and $w.$



To compute $C(v),$ we consider those spreading trees rooted at node $v.$ Given a spreading tree ${\cal P}={{\cal T}, {\bf t}}$ rooted at node $v,$ denote by $\Path(v,w)$ the set of edges on the path from node $v$ to node $w.$ The cost of the spreading tree can be written as
\begin{align}
C(\PropagationPath)&={\sum_{(h,u)\in {\cal E}({\cal T})\backslash \Path(v,w)}(\InfTimeSeqEnt_u-\InfTimeSeqEnt_h-\mu)^2 }\label{eqn: cost-1}\\
&+ {\sum_{(h,u)\in \Path(v,w)}(\InfTimeSeqEnt_u-\InfTimeSeqEnt_h-\mu)^2}\label{eqn: cost-2-nd}
\end{align}

Recall that only the infection time of nodes $v$ and $w$ are known. Furthermore, nodes $v$ and $w$ will not both appear on a path in  ${\cal T}\backslash \Path(v,w).$ Therefore, by choosing $\tau_u-\tau_h=\mu$ for each $(h,u)\in {\cal E}({\cal T})\backslash \Path(v,w),$ we have $$(\ref{eqn: cost-1})=0.$$
Next applying Lemma \ref{lem:assigntime}, we obtain that
\begin{eqnarray}
(\ref{eqn: cost-2-nd})&\geq& |\Path(v,w)|\left(\frac{\InfObsTimeEnt_w-\InfObsTimeEnt_v}{|\Path(v,w)|}-\mu\right)^2,\label{eq:thm-1}
\end{eqnarray} where the equality is achieved by assigning the timestamps according to Lemma \ref{lem:assigntime}.

For fixed $|\InfObsTimeEnt_w-\InfObsTimeEnt_v|$ and $\mu,$ we have
\begin{align*}
\frac{\partial (\ref{eq:thm-1})}{\partial |\Path(v,w)|}&=\mu^2-\left(\frac{\InfObsTimeEnt_w-\InfObsTimeEnt_v}{|\Path(v,w)|}\right)^2\\
&<_{(a)}\mu^2-\left(\frac{\mu |\InfNode|}{|\Path(v,w)|}\right)^2\\
&<_{(b)} \mu^2-\left(\frac{\mu |\InfNode|}{|\InfNode|}\right)^2=0,
\end{align*}
where inequality $(a)$ holds because of the assumption $\InfObsTimeEnt_w-\InfObsTimeEnt_v>\mu |\InfNode|$ and inequality $(b)$ is due to $|\Path(v,w)|\leq |\InfNode|-1.$ So  $(\ref{eq:thm-1})$ is a decreasing function of $|\Path(v,w)|$ (the length of the path).

Let $\LongestPathLength$ denote the length of the longest path between $v$ and $w.$ Given the longest path between $v$ and $w,$ we can construct a spreading tree ${\cal P}^*$ by generating ${\cal T}^*$ using the breadth-first search starting from the longest path and assigning timestamps ${\bf t}^*$ as mentioned above.  Then,
\begin{eqnarray}
C(v)=C({\cal P}^*)=\min_{\PropagationPath_v\in {\cal L}({\cal I}, {\bm \tau})}C(\PropagationPath_v)=\LongestPathLength\left(\frac{\InfObsTimeEnt_w-\InfObsTimeEnt_v}{\LongestPathLength}-\mu\right)^2. \label{eq:thm-case1}
\end{eqnarray}

Therefore, the algorithm that computes $C(v)$ can be used to find the longest path between nodes $v$ and $w.$ Since the longest path problem is NP-hard \cite{GarJoh_79}, the calculation of $C(v)$ must also be NP-hard.

\section*{Proof of Lemma \ref{lem:assigntime}}
\label{sec:lemma}

Define $x_{k,k-1}=t_k-t_{k-1},$ so the cost $C$ can be written as
\[
C({\bf x})=\sum_{k=2}^n(\InfTimeSeqEnt_k-\InfTimeSeqEnt_{k-1}-\mu)^2=\sum_{k=2}^n(x_{k,k-1}-\mu)^2.
\] The cost minimization problem can be written as
\begin{eqnarray}
&\min C({\bf x})=\sum_{k=2}^n(x_{k,k-1}-\mu)^2\\
\hbox{subject to:}&\sum_{k=2}^n x_{k,k-1}=t_n-t_1 \\
& x_{k,k-1}\geq 0.
\end{eqnarray}
Note that $C({\bf x})$ is a convex function in ${\bf x}.$ By verifying the KKT condition, it can be shown that the optimal solution to the problem above is $x_{k,k-1}=\frac{\tau_n-\tau_1}{n-1},$ which implies $t_k=\tau_1+(k-1)\frac{\tau_n-\tau_1}{n-1}.$

\section*{Proof of Theorem \ref{thm:complexity}}
Note that the complexity of the modified breadth first search is $O(|{\cal E}_I|)$ since each edge in the subgraph formed by the infected nodes only needs to be considered once. We next analyze the complexity of EIF:
\begin{itemize}

\item Step 1: The complexity of computing the paths from an infected node to all other infected nodes is $O(|{\cal E}_I|).$ Given $|\alpha|$ infected nodes with timestamps, the computational complexity of Step 1 is  $O(|\alpha||{\cal E}_I|).$

\item Step 2: The complexity of sorting a list of size $|\alpha|$ is $O(|\alpha|\log(|\alpha|)).$

\item Steps 3 and 4: To construct the spreading tree for a given node, $|\alpha|$ infected nodes need to be attached in Steps 3 and 4.  Each attachment requires the construction of a modified breadth-first tree, which has complexity $O(|{\cal E}_I|).$ So the overall computational complexity of Steps 3 and 4 is $O(|\alpha||{\cal E}_I|).$

\item Step 5: The breadth-first search algorithm is needed to complete the spreading tree, which has complexity $O(|{\cal E}_I|).$
\end{itemize}
From the discussion above, we can conclude that the computational complexity of constructing the spreading tree from a given node and calculating the associated cost is  $O(|\alpha||{\cal E}_I|).$  CR (or TR) repeats EIF for each infected node, with complexity $O(|\alpha||{\cal I}||{\cal E}_I|),$ and then sort the infected nodes, with complexity $O(|\cal I|\log|\cal I|).$ Therefore, the overall complexity of CR (or TR) is $O(|\alpha||{\cal I}||{\cal E}_I|).$